\documentclass[12pt]{article}
\usepackage{jheppub}
\usepackage{ifpdf}
\pdfoutput=1

\usepackage{amssymb}
\usepackage{color}
\usepackage{graphicx,psfrag, subfigure}
\usepackage{amsmath}
\usepackage{hyperref}
\usepackage{bbold}

\usepackage[applemac]{inputenc}
\usepackage{dsfont}

\usepackage{soul}
\definecolor{dm}{cmyk}{.20, 0, .30, 0}
\sethlcolor{dm}

\newcommand{\Hz}{{\cal H}_Z}

\renewcommand{\d}{\textrm{d}}

\numberwithin{equation}{section}

\def\be{\begin{equation}}
\def\ee{\end{equation}}
\def\bea{\begin{eqnarray}}
\def\eea{\end{eqnarray}}
\def\bb{\bar{b}}
\def\ba{\bar{a}}
\def\bc{\bar{c}}

\def\bW{\overline{W}}
\def\bZ{\bar Z}

\def\M{M_{Pl}}
\def\ms{m_{susy}}
\def\Ms{M^2_{Pl}}

\def\hW{\widehat{W}}
\def\Z{\hat{Z}}

\def\ho{\hat{\omega}}
\def\ha{\hat{a}}
\def\hla{\hat{\lambda}}
\def\hm{\hat{\mu}}
\def\hb{\hat{\beta}}

\newcommand{\nn}{\nonumber}
\newcommand{\lp}{\left(}
\newcommand{\rp}{\right)}
\newcommand{\ls}{\left[}
\newcommand{\rs}{\right]}

\begin{document}

\begin{titlepage}

\setcounter{page}{1} \baselineskip=15.5pt \thispagestyle{empty}

\bigskip\
\begin{center}
{\Large \bf Supersymmetric Vacua in Random Supergravity}\\
\vskip 5pt
\vskip 15pt
\end{center}
\vspace{0.5cm}
\begin{center}
{
Thomas C. Bachlechner, David Marsh, Liam McAllister, and Timm Wrase}
\end{center}

\vspace{0.05cm}

\begin{center}
\vskip 4pt
\textsl{Department of Physics, Cornell University,
Ithaca, NY 14853 USA}

\end{center} 
{\small  \noindent  \\[0.2cm]
\noindent

We determine the spectrum of scalar masses in a supersymmetric vacuum of a general ${\cal N}=1$ supergravity theory, with the Kähler potential and superpotential taken to be random functions of $N$ complex scalar fields. We derive a random matrix model for the Hessian matrix and compute the eigenvalue spectrum. Tachyons consistent with the Breitenlohner-Freedman bound are generically present, and although these tachyons cannot destabilize the supersymmetric vacuum, they do influence the likelihood of the existence of an `uplift' to a metastable vacuum with positive cosmological constant. We show that the probability that a supersymmetric AdS vacuum has no tachyons is formally equivalent to the probability of a large fluctuation of the smallest eigenvalue of a certain real Wishart matrix. For normally-distributed matrix entries and any $N$, this probability is given exactly by $P = {\rm exp}(-2N^2|W|^2 /\ms^{2})$, with $W$ denoting the superpotential and $\ms$ the supersymmetric mass scale; for more general distributions of the entries, our result  is accurate when $N \gg 1$. We conclude that for $|W| \gtrsim m_{susy}/N$, tachyonic instabilities are ubiquitous in configurations obtained by uplifting supersymmetric vacua.
}

\vspace{0.3cm}

\vspace{0.6cm}

\vfil
\begin{flushleft}
\small \today
\end{flushleft}
\end{titlepage}

\section{Introduction}\label{intro}

Supersymmetric vacua of four-dimensional supergravity provide a computable stepping stone toward realistic non-supersymmetric vacua of string theory.
Supersymmetry secures vacuum stability, and also opens powerful analytic perspectives, such as the $AdS_4/CFT_3$ correspondence, that are unavailable in de Sitter solutions.  AdS vacua have played a prominent role in the development of mechanisms for moduli stabilization, as exemplified by~\cite{KKLT}: a supersymmetric AdS vacuum serves as a controllable foundation, and a source of supersymmetry breaking is introduced to lift the energy density to positive values.
However, tachyons are abundant in supersymmetric AdS vacua, and may pose significant obstacles to uplifting AdS solutions to metastable de Sitter vacua.

In this work we use random matrix theory to obtain the scalar mass spectrum in a supersymmetric vacuum of a generic four-dimensional ${\cal N}=1$ supergravity.
We take the Kähler potential $K$ and superpotential $W$ to be random functions of $N$ complex scalar fields,  in a sense made precise in \cite{Marsh:2011aa}.
The Hessian matrix ${\cal H}$ is a particular combination of the derivatives of $W$ and $K$,
and the techniques of random matrix theory can be used to determine the spectrum of eigenvalues of ${\cal H}$, i.e.\ the scalar mass spectrum. Strong correlations among the eigenvalues make large fluctuations --- for example, a fluctuation resulting in a tachyon-free spectrum --- extremely unlikely at large $N$.

The key results of this work are statements about the prevalence of tachyons in supersymmetric AdS vacua.
AdS supersymmetry (see \cite{deWit:1999ui} for a review) ensures that the real and imaginary parts of each complex scalar field have distinct masses \cite{Burges:1985qq}, with splitting proportional to the AdS scale, set by $|W|$.  The mass spectrum therefore consists of two branches.
We show that the spectrum of each branch is determined by the eigenvalue spectrum of a particular ensemble of real Wishart matrices, and we give an analytic result for the full spectrum.
The mass spectrum critically depends on the ratio of the AdS scale to the supersymmetric mass scale, denoted $m_{susy}$: for $m_{susy}/|W| \to \infty$ all scalars have positive mass-squared, with spectrum given by the Mar\v{c}enko-Pastur law \cite{MR0208649}, while for $m_{susy}/|W| \to 0$  tachyons are endemic and the spectrum asymptotes to that of the Altland-Zirnbauer C$I$ ensemble \cite{Altland:1997zz}.

Using a result of Edelman \cite{Edelman:1988:ECN:58846.58854}, we obtain the probability $P$ of a fluctuation that renders all scalars non-tachyonic, in terms of the probability of a corresponding fluctuation in the associated Wishart ensemble: we find that $P = {\rm exp}(-2N^2|W|^2 /\ms^{2})$.  Then, building on results from the Coulomb gas  formulation of random matrix theory, we obtain the mass spectrum resulting from such a fluctuation. We conclude that AdS vacua with  $m_{susy}^2 \ll N^2 |W|^2$
are overwhelmingly likely to contain tachyons allowed by the Breitenlohner-Freedman (BF) bound.
Finally, we determine the probability that a de Sitter critical point obtained by uplifting a supersymmetric AdS vacuum is metastable, for a range of uplifting scenarios (see also \cite{Chen:2011ac}).\\
\indent The organization of this paper is as follows.
In \S\ref{preliminaries} we present the structure of the Hessian matrix in a supersymmetric AdS vacuum.
After briefly reviewing the Wishart ensemble \cite{1928} and the Altland-Zirnbauer C$I$
ensemble \cite{Altland:1997zz}, we show that the eigenvalue spectrum  of the Hessian matrix is determined by the spectrum of the Altland-Zirnbauer C$I$
ensemble, which in turn can be related to the spectrum of a particular ensemble of Wishart matrices.  In \S\ref{spectrum} we compute the spectrum of the Hessian matrix and determine the probability of fluctuations of its smallest eigenvalue.  We then study the distribution of tachyon-free supersymmetric vacua and argue that a fraction $\propto \frac{1}{N^2}$ of the parameter space contains no tachyons at all.
In \S\ref{implications} we describe the implications of our results, focusing on instabilities in uplifted vacua.  We conclude in \S\ref{conclusions}.
In the appendix we present the results of extensive numerical simulations that confirm our analytic findings.

Unless otherwise specified, we work in natural units with $\M^{-2}=8\pi G_N =1$.

\section{The Hessian Matrix in a Supersymmetric Vacuum} \label{preliminaries}

We begin in \S\ref{structure} by presenting the structure of the Hessian matrix in a supersymmetric vacuum.
Then, in \S\ref{ensemblereview}, we briefly recall two relevant ensembles of random matrices, the Wishart ensemble and the Altland-Zirnbauer C$I$ ensemble,
and we express the eigenvalues of the Hessian matrix in terms of those of the Wishart and Altland-Zirnbauer C$I$ ensembles.

\subsection{Supersymmetric vacua in ${\cal{N}}=1$ supergravity} \label{structure}

Supersymmetric points satisfying $F_a = D_a W \equiv (\partial_a + \partial_a K)W  =0$ are critical points of the ${\cal N} =1$ supergravity F-term potential, $V_F = e^K \left( F_a \bar F^a - 3 |W|^2 \right)$.
The Hessian matrix at such a point is given by
\bea
{\cal H} &=&
\left(
\begin{array}{c c}
\partial^2_{a \bb} V & \partial^2_{a b} V \\
\partial^2_{\ba \bb} V & \partial^2_{\ba b} V
\end{array}
\right)
=
{\cal H}_Z
 - 2 |W|^2 \mathbb{1}  \,  , \label{eq:Hess}
\eea
where
\bea
{\cal H}_Z =
\left(
\begin{array}{c c}
 Z_{a}^{~\bar c}\ \bZ_{\bb \bc}  &  - Z_{a b} \bW \\
 - \bZ_{\ba \bb} W & \bZ_{\ba}^{~c}\ Z_{b c}
\end{array}
\right)  \, , \label{Q}
\eea
and $Z_{ab} = Z_{ba}\equiv {\cal D}_a D_b W$ is the supersymmetric fermion mass matrix.  Here ${\cal D}_a$ denotes the geometrically and  Kähler covariant derivative, and we
have specialized to a Kähler gauge in which $\langle K \rangle = 0$ at the critical point.
The effect of the contribution proportional to the unit matrix is to shift the entire spectrum to more negative values by an amount $2|W|^2$. Therefore, finding the spectrum of the
Hessian in a supersymmetric vacuum amounts to determining the spectrum of the matrix ${\cal H}_Z$, which will be the focus of much of this work.

To simplify ${\cal H}_Z$, we first write $Z = U \Sigma U^T$, where $U$ is a unitary matrix whose columns are orthonormal eigenvectors of $Z\bZ$, $\Sigma = {\rm diag}(\lambda_1,\ldots \lambda_N)$, and the $\lambda_a$ are real and nonnegative, with $\lambda_a^2$ the eigenvalues of $Z\bZ$  \cite{HornAndJohnson}.
This Takagi factorization exists for any complex symmetric matrix $Z$.
Performing the $2N \times 2N$ unitary transformation
\bea
{\cal H}_Z \to {\cal U}^{\dagger} {\cal H}_Z\  {\cal U} \quad \text{with} \quad
{\cal U} =
\left(
\begin{array}{c c}
U  &  0\\
0 & U^*
\end{array}
\right) , \label{U}
\eea
the matrix ${\cal H}_Z$ can be written as
\bea
{\cal H}_Z =
\left(
\begin{array}{c c}
\Sigma^2 &  - \Sigma \bW \\
 - \Sigma W & \Sigma^2 \end{array}
\right)  \, . \label{D}
\eea
After rearranging the rows and columns in an obvious way, ${\cal H}_Z$ takes the block diagonal form
\bea
{\cal H}_Z &=&
\left(
\begin{array}{c c c c c}
\lambda_1^2 & - \bW \lambda_1 & 0 &0 \\
- W \lambda_1 & \lambda_1^2 & 0& 0 \\
0 & 0 & \lambda_2^2 & - \bW \lambda_2  \\
0 & 0& - W \lambda_2 & \lambda_2^2 \\
&&&& \ddots
\end{array} \right)  \, . \label{Hess}
\eea
The spectrum of each block of ${\cal H}_Z$ is
\be
\omega_{a \pm} \equiv \lambda_a^2 \pm |W| \lambda_a \, . \label{eq:omega}
\ee
Thus, the scalar mass spectrum in a supersymmetric vacuum is fully determined by the vev of the superpotential, $W$, and by the spectrum of eigenvalues $\lambda_a^2$ of  $Z\bZ$.

The eigenvalue $\omega_{a -}$ is minimized by  $\lambda_a = |W|/2$, which gives   $\omega_{a -}=-|W|^2/4$. So the smallest possible eigenvalue of ${\cal H}$ is
\be
m_{\text{min}}^2 = -\frac{|W|^2}{4} - 2 |W|^2 = -\frac94 |W|^2 = -\frac34 |V| = m_{\text{BF}}^2\,, \label{eq:BF}
\ee
where we have used that $V = -3|W|^2$ at a supersymmetric minimum, and $m_{\text{BF}}$
denotes the Breitenlohner-Freedman bound \cite{Breitenlohner:1982bm, Breitenlohner:1982jf}.

\subsection{Classical ensembles} \label{ensemblereview}
In \S\ref{sec:spectrum}, we will obtain the spectrum of the Hessian matrix using the methods of random matrix theory.  In this section, we review the essential properties of the two matrix ensembles that are relevant for the analysis of \S\ref{sec:spectrum}: the Wishart ensemble and the Altland-Zirnbauer C$I$ ensemble.

\subsubsection{The Wishart ensemble}\label{sec:Wishart}

The ensemble of {\it{Wishart matrices}} ${\cal W}$ \cite{1928}  takes the form
\begin{equation} \label{Wishartdef}
{\cal W}=A A^{\dagger}\,,
\end{equation}
where $A$ is an $N \times M$ real or complex matrix whose entries are independent and identically distributed (i.i.d.)\ random variables drawn from a statistical distribution with mean zero and variance $\sigma^2$, which we denote by $\Omega(0,\sigma)$.
The probability density of the eigenvalues of ${\cal W}$ depends on the ratio $M/N$, and for our purposes it will suffice to consider the case $M \ge N$.

Since a Wishart matrix is the Hermitian square of another matrix, it is necessarily positive semidefinite.  Upon changing variables to an eigenbasis of ${\cal W}$, the Jacobian of the transformation induces `interaction terms' between the different eigenvalues $\mu_a$.
For entries drawn from a normal distribution,\footnote{Universality implies that more general distributions yield equivalent results at sufficiently large $N$ \cite{2006math.ph...3038D, 2007JPhA...40.4317V, 2011arXiv11035922K, TaoVu, Erdos}, which we have verified directly in our simulations.}
the joint probability density is (cf.\ \cite{CambridgeJournals:298726})
\be
f(\mu_1, \ldots, \mu_N) = {\cal C}\ \exp\lp- \frac{\beta}{2} \lp \frac{1}{\sigma^2} \sum_{a=1}^N \mu_a \ - 2\sum_{a < b}^N{\rm{ln}}|\mu_a - \mu_b| - \xi\,\sum_{a=1}^N {\rm{ln}}\,\mu_a \rp\rp \, , \label{eq:Wishart}
\ee
where $\xi= M-N+1-2/\beta$, and $\beta=1,2$ for real and complex matrices, respectively.
In the Coulomb gas picture in which  the joint probability density is interpreted as an exponential of the free energy of an ensemble of $N$ interacting particles, the non-negativity of a Wishart matrix corresponds to the presence of a hard wall at $\mu=0$.

The probability density function (pdf) for the eigenvalues of ${\cal W}$  is given by the
Mar\v{c}enko-Pastur law \cite{MR0208649},
\begin{equation} \label{eq:wishart}
\rho_{MP}(\mu) =  \frac{1}{2\pi N \sigma^2 \mu}\sqrt{(\eta_+-\mu)(\mu-\eta_-)} \,,
\end{equation}
where
\begin{equation} \label{etadefinition}
\eta_\pm = N \sigma^2 (1\pm\sqrt{\eta})^2\,,
\end{equation}
and $\eta=M/N \geq 1$.

\begin{figure}
  \begin{center}
    \includegraphics[width=0.6\textwidth]{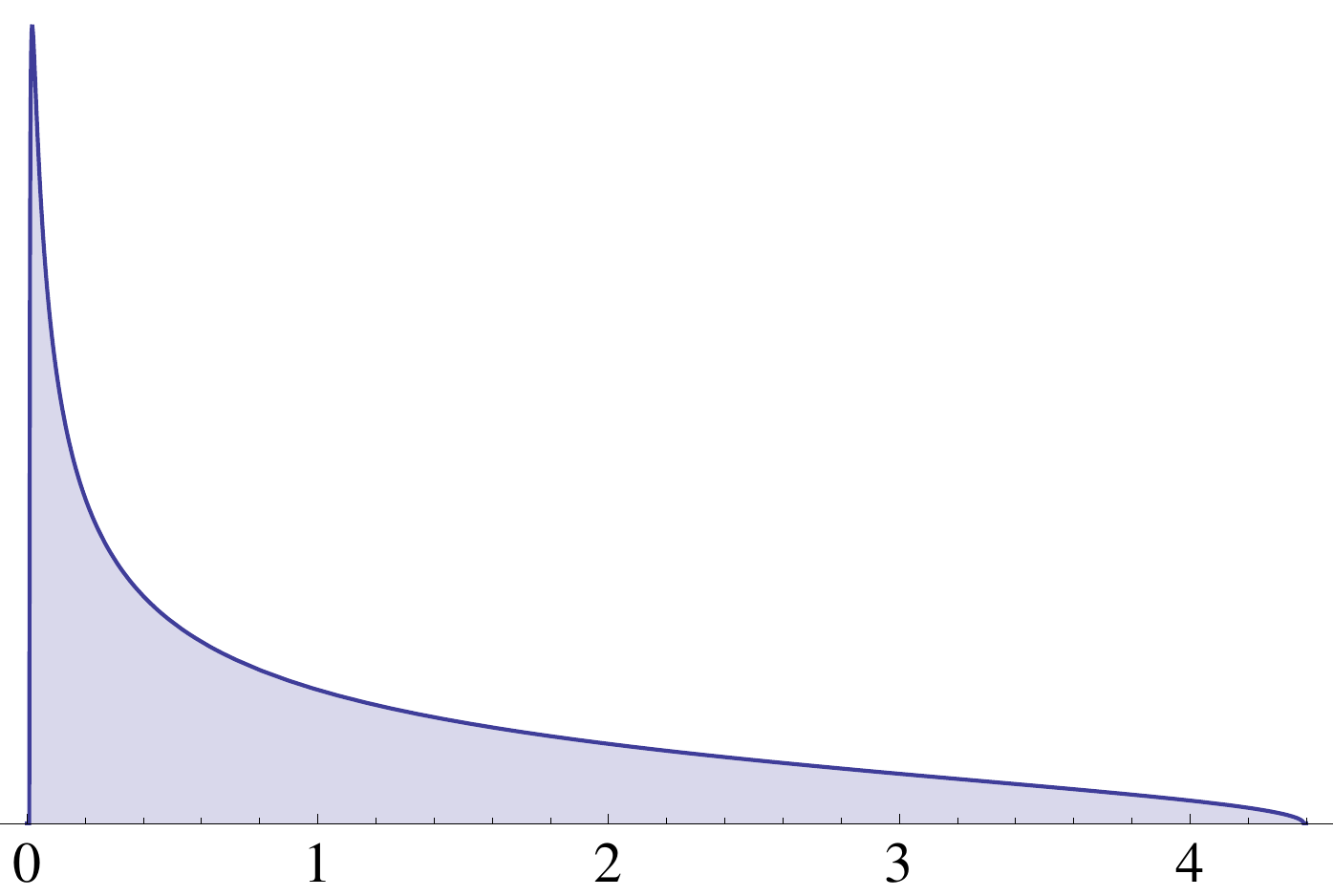}
  \end{center}
  \caption{The eigenvalue spectrum $\rho_{MP}(\mu)$ for an almost square Wishart matrix with $M=N+1=6$ and $\sigma=1/\sqrt{N}$.}\label{fig:Wishart}
\end{figure}

In \S\ref{sec:spectrum}, we will be led to focus on a specific ensemble of Wishart matrices in which $M = N+1$, so that $A$ is `almost square'.  By an abuse of language, we will likewise refer to the associated Wishart matrices ${\cal W}$ as `almost square', though of course ${\cal W}$ is square for any $M,N$.
For this ensemble,
$\eta = 1 + 1/N$, so that to lowest order in a $1/N$ expansion, with $\sigma = 1/\sqrt{N}$, we have $\eta_{-} \approx 0$ and $\eta_{+} \approx 4$.
A plot of the eigenvalue spectrum for this almost square case is shown in Figure \ref{fig:Wishart}.

\subsubsection{The Altland-Zirnbauer C$I$ ensemble}
\label{sec:CI}
The Altland-Zirnbauer C$I$ ensemble \cite{Altland:1997zz} consists of matrices of the form
\be\label{eq:M}
{\cal M} = \left(
      \begin{array}{cc}
        0 & Z_{a b} \\
        \bar{Z}_{\bar a  \bar b} & 0 \\
      \end{array}
    \right) \,,
\ee
where $Z_{ab}$ is a complex symmetric matrix with independent entries distributed as
\begin{equation} \label{zdistributor}
Z_{ab} \in \Omega(0, \sigma)  \quad \text{for} \quad  a\neq b  \quad \text{and} \quad  Z_{aa} \in \Omega(0, \sqrt{2} \sigma)  \quad \text{(no sum on $a$)}\,.
\end{equation}
Equivalently, the Altland-Zirnbauer C$I$ ensemble can be defined as
the matrix ensemble satisfying
\bea
T {\cal M} T^{\dagger} &=& {\cal M}^* \, ,  \label{eq:Tsymm} \\
C {\cal M} C^{\dagger} &=& - {\cal M} \, , \\
P {\cal M} P^{\dagger} &=& - {\cal M} \, ,
\eea
for some unitary transformations $C, P,$ and $T$ satisfying $T^\top = + T$, $C^\top  = - C$ and $P^\top  = + P$, acting on a Hermitian matrix ${\cal M}$. For C$I$ representations of the form \eqref{eq:M}, we identify $C = \sigma_2$, $P=\sigma_3$, and $T= -i\sigma_1$, where $\sigma_i$ denote the Pauli matrices that act on the $N\times N$ block matrices of ${\cal M}$.

\begin{figure}
  \begin{center}
    \includegraphics[width=0.6\textwidth]{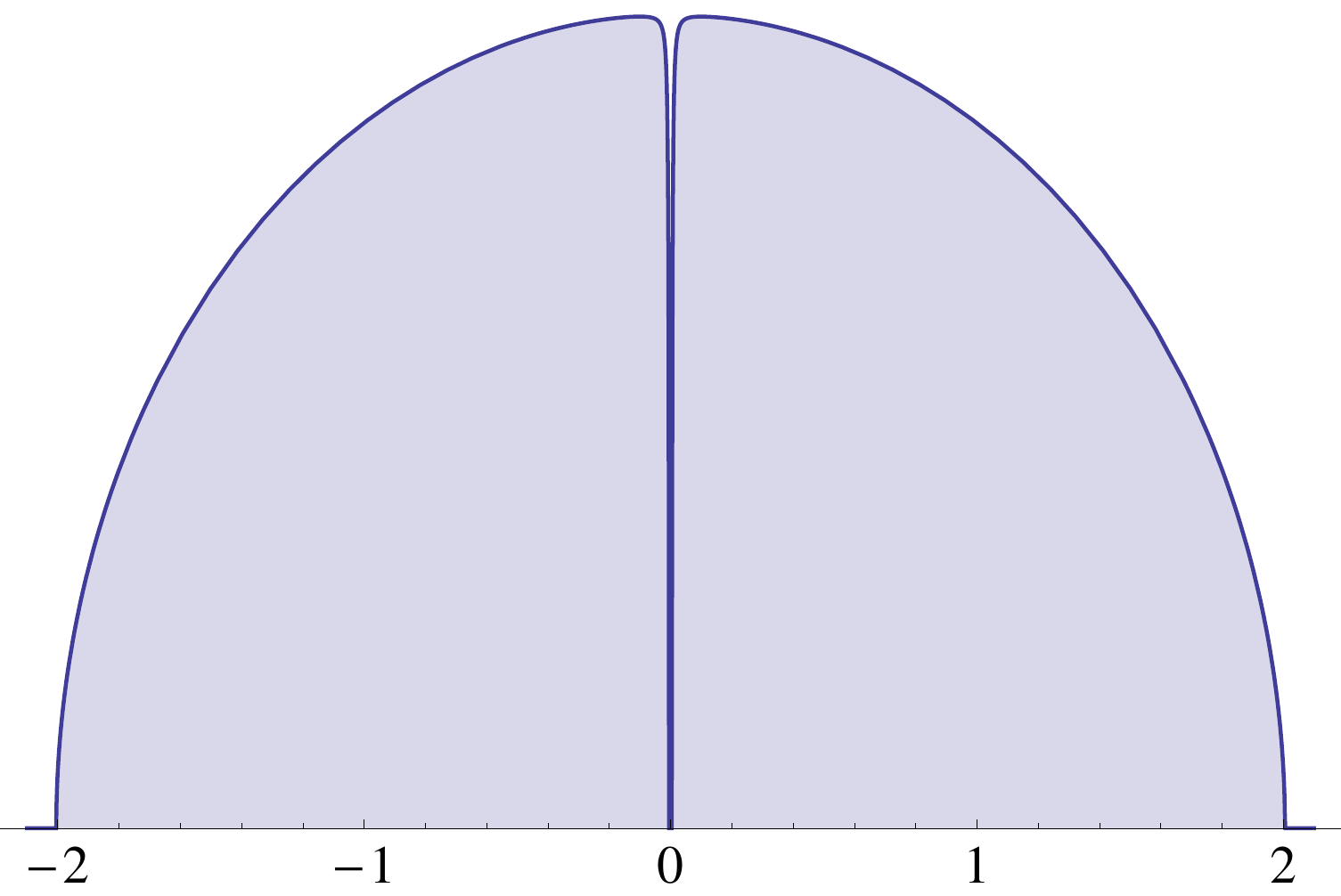}
  \end{center}
  \caption{The eigenvalue spectrum of the Altland-Zirnbauer C$I$ ensemble for $N=100$.}\label{fig:AZCI}
\end{figure}

The eigenvalue spectrum of ${\cal M}$ is reminiscent of the Wigner semicircle law, but the $2N$ eigenvalues of ${\cal M}$ come in opposite-sign pairs $\pm \nu_a$, with  $0\le \nu_1 \le \ldots \le \nu_{N}$.
Taking $\Omega(0, \sigma)$ to be a normal distribution,
the joint probability density of the $N$ positive eigenvalues  is given by
\be
f(\nu_1,\ldots,\nu_N) = {\cal{C}}\ {\rm{exp}}\lp-\frac{1}{2 \sigma^2} \sum_{a=1}^{N} \nu_{a}^2+  \sum^N_{a< b}{\rm{ln}}|\nu^2_a-\nu^2_b| + \sum^N_{a=1} \ln  \nu_a \rp\,. \label{eq:CI}
\ee
In the Coulomb gas picture, the term $\sum^N_{a=1} \ln \nu_a$ encodes a repulsive force between each mirror pair of eigenvalues, $\pm\nu_a$. In the eigenvalue spectrum this leads to a cleft at the origin, which is a subleading $1/N$ effect. The eigenvalue spectrum is shown in Figure \ref{fig:AZCI}.

The Altland-Zirnbauer C$I$ ensemble has played a prominent role in the study of critical points in supergravity  \cite{Denef:2004cf}: the critical point equation $\partial_{a} V_F = 0$ may be written as an eigenvalue equation for a matrix ${\cal M}$ of the form (\ref{eq:M}),\footnote{Our equation \eqref{eq:M} differs from the definition of ${\cal M}$ in \cite{Denef:2004cf} by the phase of $W$, cf.~equation \eqref{eq:Mtilde}, but the distinction is unimportant here.}
where $Z_{ab} = Z_{ba}\equiv {\cal D}_a D_b W$ is the fermion mass matrix. From studies of type IIB flux vacua \cite{Denef:2004ze}, it is well-motivated to take the distribution (\ref{zdistributor}) of the entries of $Z_{ab}$, i.e.~$\Omega(0, \sigma)$, to be uniform with support up to the flux scale. However, by universality  only the first few moments  of the distribution $\Omega(0, \sigma)$ are important at large $N$ \cite{2006math.ph...3038D, 2007JPhA...40.4317V, 2011arXiv11035922K, TaoVu, Erdos}, and the exact distribution of the entries of $Z_{ab}$ is therefore of little significance.

With this identification of $Z_{ab}$,
we immediately see that the positive eigenvalues $\nu_a$ of ${\cal M}$ are identical to the diagonal entries $\lambda_a$ of the Takagi factorization of  $Z$ presented in \S\ref{structure}.
Thus, the scalar mass spectrum in a supersymmetric vacuum can be expressed in terms of the eigenvalue spectrum of the Altland-Zirnbauer C$I$ ensemble.
We will find it useful to relate the latter to the spectrum of a certain Wishart ensemble, as we now explain.

\subsubsection{The relation between the C$I$ ensemble and the Wishart ensemble} \label{sec:WishCI}

To conclude this review, we will point out a formal equivalence between the squares of the eigenvalues of the C$I$ distribution, and the eigenvalues of an almost square real Wishart matrix.
The joint probability measure of the C$I$ ensemble \eqref{eq:CI} is
\be
f(\nu_1,\ldots,\nu_N)\, \d \nu_1 \ldots \d \nu_N =  2^{-N}{\cal{C}}\ {\rm{exp}}\Bigl(-\frac{1}{2 \sigma^2} \sum_{a=1}^{N} \nu_{a}^2+  \sum^N_{a< b}{\rm{ln}}|\nu^2_a-\nu^2_b|\Bigr)  \prod^N_{a=1} \d \nu_a^2  \, .
\ee
Defining $\mu_a = \nu_a^2$ and $ \tilde{{\cal{C}}} = 2^{-N}{\cal{C}}$, this takes the form
\be
f(\mu_1,\ldots,\mu_N)\, \d \mu_1 \ldots \d \mu_N = \tilde{{\cal{C}}}\ {\rm{exp}}\Bigl(-\frac{1}{2 \sigma^2} \sum_{a=1}^{N} \mu_{a}+  \sum^N_{a< b}{\rm{ln}}|\mu_a-\mu_b|\Bigr)  \prod^N_{a=1} \d \mu_a \,, \label{eq:Jointpdf}
\ee
which is the joint probability measure for the Wishart ensemble (cf.~equation \eqref{eq:Wishart}) with $\beta=1$ and $\xi=0$. Since  $\xi = M-N +1 - 2/\beta$, this implies that $M=N+1$, so that  the distribution of the squares of the eigenvalues $\nu$ of the C$I$ ensemble is equivalent to the distribution of the eigenvalues $\mu$ of the ensemble of real, almost square Wishart matrices.

A further clarification is appropriate.  The matrix $Z\bZ$ is the Hermitian square of the square matrix $Z$, so one might incorrectly suppose that with the entries of $Z$ distributed according to (\ref{zdistributor}), the ensemble of matrices $Z\bZ$ corresponds to the complex Wishart ensemble with $M=N$.  However, $Z$ is symmetric, while the matrix $A$ appearing in the defining relation (\ref{Wishartdef}) has no special symmetry properties.  What we have just seen is that in fact the ensemble of matrices $Z\bZ$ corresponds to the {\it{real}} Wishart ensemble with $M=N+1$.

\section{The Spectrum of the Hessian Matrix} \label{spectrum}
\label{sec:spectrum}

In the preceding section we showed that the scalar mass spectrum in a supersymmetric vacuum is completely determined (for fixed vev $|W|$) by the eigenvalue spectrum of the Altland-Zirnbauer C$I$ ensemble, which in turn can be obtained from the spectrum of the real Wishart ensemble with $M=N+1$. In this section we will use existing results for the Wishart ensemble to obtain an analytic expression for the scalar mass spectrum.

In \S\ref{sec:Hspectrum} we will show that an  AdS vacuum with $N \gg1$ and $W \approx Z_{ab}$ has a large number of BF-allowed tachyons, which may be problematic for model building. In \S\ref{sec:fluctuations} we compute the fraction of vacua in which there are no tachyons, and we obtain the mass spectrum in these rare vacua.

\subsection{The spectrum of ${\cal H}_Z$ from the Mar\v{c}enko-Pastur law} \label{sec:Hspectrum}

From (\ref{eq:omega}) we note that the spectrum of ${\cal H}_Z$
interpolates between that of  Mar\v{c}enko-Pastur (for small vevs of $|W|$), and that of the C$I$ ensemble (for $|W| \rightarrow \infty$).
More precisely, this interpolation is controlled by the dimensionless ratio $\ms/m_{3/2}$, where $\ms$ is the supersymmetric mass scale, i.e.\ the scale of the entries of $Z_{ab}$, and\footnote{Recall that in each vacuum the expectation value of the Kähler potential has been set to zero.} $m_{3/2} = |W|/\Ms$.
To be fully explicit, we will temporarily reinstate factors of $\M$, and we factor out a scale $\ms$ from the tensor $Z_{ab}$ as
\be\label{eq:Zhat}
Z_{ab} = \ms~\Z_{ab} \, ,
\ee
where $\ms$ is determined so that\footnote{Choosing the standard deviation of $\Z_{ab}$ to be proportional to $1/\sqrt{N}$ yields mass spectra that are $N$-independent to leading order in $1/N$; the standard deviation of the $Z_{ab}$ can then be changed by rescaling $\ms$.} $\Z_{ab}= \Z_{ba} \in \Omega(0,1/\sqrt{N})$, for $a \neq b$, and $\Z_{aa} \in \Omega(0,\sqrt{2/N})$.
Finally, we introduce the notation
\be
 |W| = m_{3/2}~\Ms = \ms~\hW~\Ms \, ,
\ee
for a dimensionless parameter $\hW$. The eigenvalues of the  matrix $\widehat{M} = {\cal M}/\ms$ are correspondingly denoted by $\hla$, and we note that the  spectrum of each $2\times2$ block of $\Hz$ can be written as (by here and henceforth suppressing indices)
\be
 \omega_{\pm} \equiv \ms^2~\ho_{\pm}   = \ms^2\left(\hla^2 \pm |\hW| \hla  \right) \, , \label{eq:omega2}
\ee
with $\hla \geq 0$.  We will now obtain the distribution of $\omega_{\pm}$
for arbitrary values of $|\hW|$.

The probability that $\ho_{\pm}$ is smaller than a given value $\ha$ is given by
\bea
&P\lp \ho_{\pm} \leq \ha \in {\cal D}_{\pm} \rp = P\lp\hla^2 \pm |\hW| \hla  \leq \ha \rp  =   P\left(  \left|\hla \pm \frac{|\hW|}{2} \right|  \leq \sqrt{\ha + \frac{|\hW|^2}{4}} \right) &\nn \\
&= P\left(\hla   \leq \mp \frac{|\hW|}{2} + \sqrt{\ha + \frac{|\hW|^2}{4}} \right) - P\left(\hla  \leq \mp \frac{|\hW|}{2} -\sqrt{\ha + \frac{|\hW|^2}{4}} \right) \label{eq:PHz}
\, ,&
\eea
where the domains of support for the probability density functions of the two branches of ${\cal H}_Z$ are given by ${\cal D}_+ = [0, \infty)$ and ${\cal D}_- = [-|\hW|^2/4, \infty)$, respectively. Due to the positivity of $\hla$, the last term of \eqref{eq:PHz} does not contribute to the probability of the positive definite branch, and it only contributes to the $\ho_-$ branch for $\ha<0$.

Equation \eqref{eq:PHz} demonstrates how the probability density of the eigenvalues of the supersymmetric Hessian matrix is completely specified by the distribution of $\hla$, which is given by the  Altland-Zirnbauer C$I$ ensemble. But as shown in \S\ref{sec:WishCI}, the joint probability density of the \emph{squares} of the eigenvalues in the  C$I$  ensemble is equivalent to that of the eigenvalues of `almost square' ($M = N+1$), real Wishart matrices, as in equation \eqref{eq:Jointpdf}. Thus, the probability distribution of the eigenvalues of the Hessian matrix may be written in terms of the Wishart eigenvalue $\hm \equiv \hla^2$ as
\be
P\lp\ho_{+} \leq \ha \in {\cal D}_+\rp =  P\left(\hm \leq  \hb_- \right) \, , \label{eq:Oplus}
\ee
where $\hb_{-} = \frac{|\hW|^2}{2}+\ha -|\hW| \sqrt{\ha + \frac{|\hW|^2}{4}}$, and
\bea
&&P(\ho_{-} \leq \ha \in {\cal D}_-)
 = P\left(\hm  \leq \hb_+ \right)
-  P\left(\hm  \leq \hb_-\Big| \ha<0\right)  \, , \label{eq:Ominus}
\eea
with $\hb_{+} = \frac{|\hW|^2}{2}+\ha +|\hW| \sqrt{\ha + \frac{|\hW|^2}{4}}$. The probability density of the two branches of eigenvalues of ${\cal H}_Z$ is then given by
\bea
 \rho_{\omega_{+}}(\ha) =\frac{d P(\ho_{+} \leq \ha )}{d\ha} = \frac{d \hb_-}{d \ha}~\frac{dP(\hm \leq \hb_-)}{d \hb_-}
 = \frac{d \hb_-}{d \ha}~\rho_{MP}(\hb_-)
  \, ,
  \eea
 for $\ha \in {\cal D}_+$, and
  \bea
  \rho_{\omega_{-}}(\ha) &=& \frac{d P(\ho_{-} \leq \ha )}{d\ha} = \frac{d \hb_+}{d\ha}~\frac{dP(\hm \leq \hb_+)}{d \hb_+} - \frac{d \hb_-}{d \ha}~\frac{dP(\hm \leq \hb_- | \ha<0)}{d \hb_-} \nn \\
&=&  \frac{d \hb_+}{d \ha}~\rho_{MP}(\hb_+) - \Theta(- \ha)~\frac{d \hb_-}{d \ha}~\rho_{MP}(\hb_-) \, ,
\eea
for $\ha \in {\cal D}_-$, where $\rho_{MP}(\hb_\pm)$ is given by \eqref{eq:wishart} for $M=N+1$. The spectrum of the ensemble of  ${\cal H}_Z$ matrices expressed in terms of $\ha$ (i.e.~in units of the supersymmetric mass scale) is then given by
\bea
\rho_{{\cal H}_Z}(\ha) &=& \frac{1}{2} \left(      \rho_{\omega_{+}}(\ha) + \rho_{\omega_{-}}(\ha) \right)  \nn  \\
&=& \frac{1}{2} \left(   \frac{d \hb_+}{d \ha}~\rho_{MP}(\hb_+) +  \Theta(\ha)~\frac{d \hb_-}{d \ha}~\rho_{MP}(\hb_-) -   \Theta(-\ha)~\frac{d \hb_-}{d\ha}~\rho_{MP}(\hb_-)  \right) \nn \\
&=&  \frac{1}{2} \Biggl[ \lp 1 + \tfrac{|\widehat W|}{\sqrt{4 \ha +|\widehat W|^2}} \rp \rho_{MP}(\hb_+) + \text{sgn}(\ha) \lp 1 - \tfrac{|\widehat W|}{\sqrt{4\ha +|\widehat W|^2}} \rp \rho_{MP}(\hb_-)  \Biggr], \label{eq:omegampdf}
\eea
which has support in the domain $\ha \in {\cal D}_- = [- |\hW|^2/4, \infty)$.  Upon including the shift given in equation \eqref{Hess}, the probability density of the scalar mass spectrum is
\be
\rho_{\cal H}( m^2) =  \rho_{{\cal H}_Z}\left(m^2 + 2|W|^2\right)\,,
\ee  where $\rho_{{\cal H}_Z}$ is given in equation (\ref{eq:omegampdf}).
This is one of our principal results.

Some features of this distribution of supersymmetric masses are worth extra attention. First, we have shown that the spectrum of the Hessian matrix is given in terms of a certain Mar\v{c}enko-Pastur law, and features of the Wishart spectrum give rise to characteristic features in the spectrum of ${\cal H}_Z$ and of the Hessian ${\cal H}$.

Expanding the edge positions $\eta_{\pm}$ given in equation (\ref{etadefinition}) to leading order in $1/N$, $\rho_{MP}(\hm)$ of equation \eqref{eq:wishart} for $M=N+1$ is nonvanishing only between the `hard edge' at $\hm = 0$ and the `soft edge' at  $\hm = 4$. Since $\ho_{\pm} = \hm \pm |\hW| \sqrt{\hm}$, this means that the positive branch of ${\cal H}_Z$ has support for
\be
0 \leq \omega_{+} \leq 4 \ms^2+2|W| \ms \, ,
\ee
to leading order in $1/N$. The domain of support of the negative branch depends on $|W|$: for $|W| > 4\ms$, the probability density function is nonvanishing for
\be
4\ms^2-2|W| \ms \leq \omega_{-} \leq 0 \, ,
\ee
while for $|W| \leq 4~\ms$, the spectrum develops a hard edge at $\omega_- =-|W|^2/4$, and has support in the range
\be
-\frac{|W|^2}{4} \leq \omega_{-} \leq \text{Max}(0,4\ms^2-2|W| \ms) \, . \label{eq:omegamspectrum}
\ee
In particular, for $|W|\geq 2~m_{susy}$ the domains of support of the two branches do not overlap.

\begin{figure}[ht!]
    \begin{center}
     \subfigure[$|W|=\tfrac{1}{100}~\ms$]{\label{fig:first}
        \includegraphics[width=0.43\textwidth]{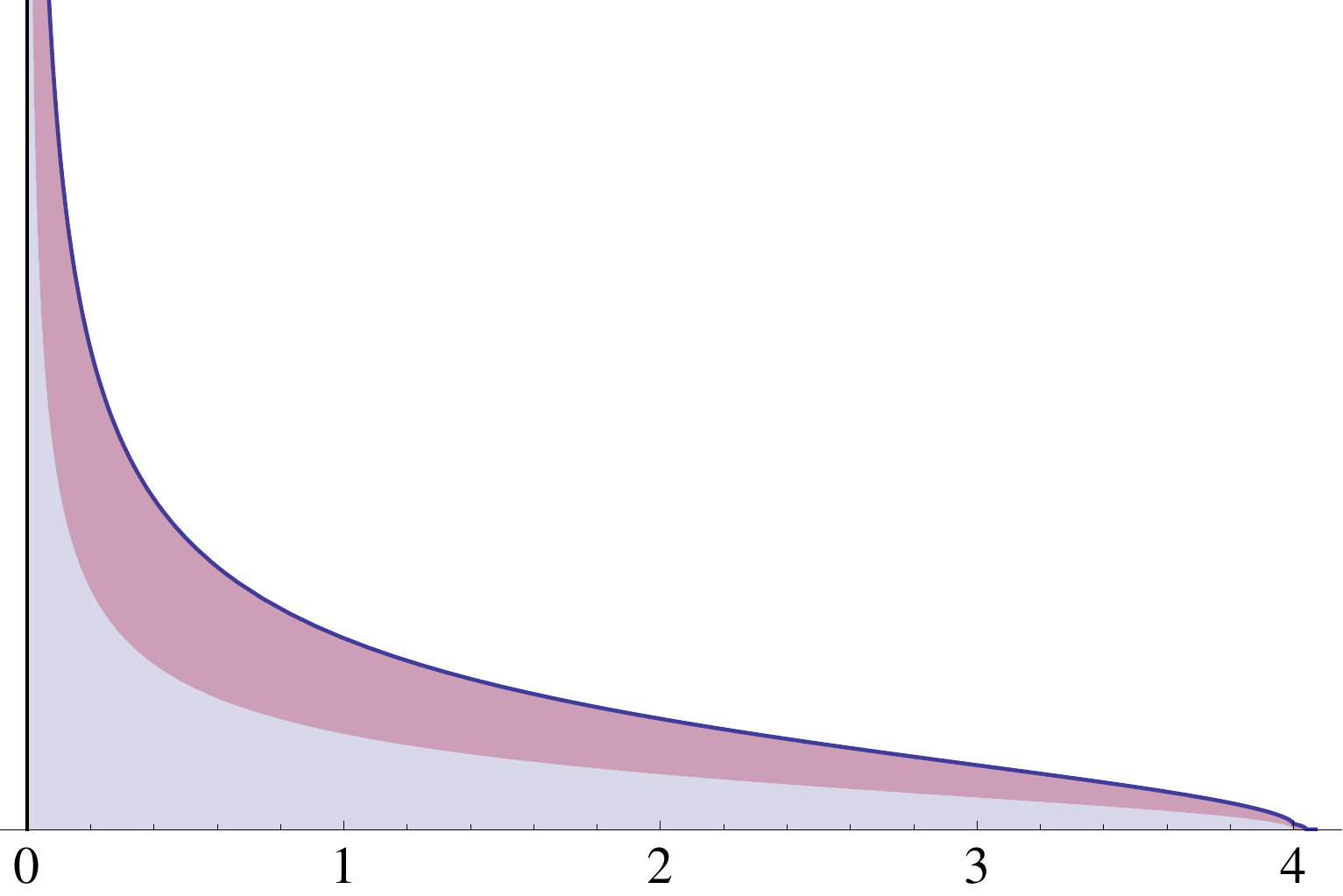}}
     \subfigure[$|W|=\tfrac{1}{2}~\ms$] {
        \includegraphics[width=0.43\textwidth]{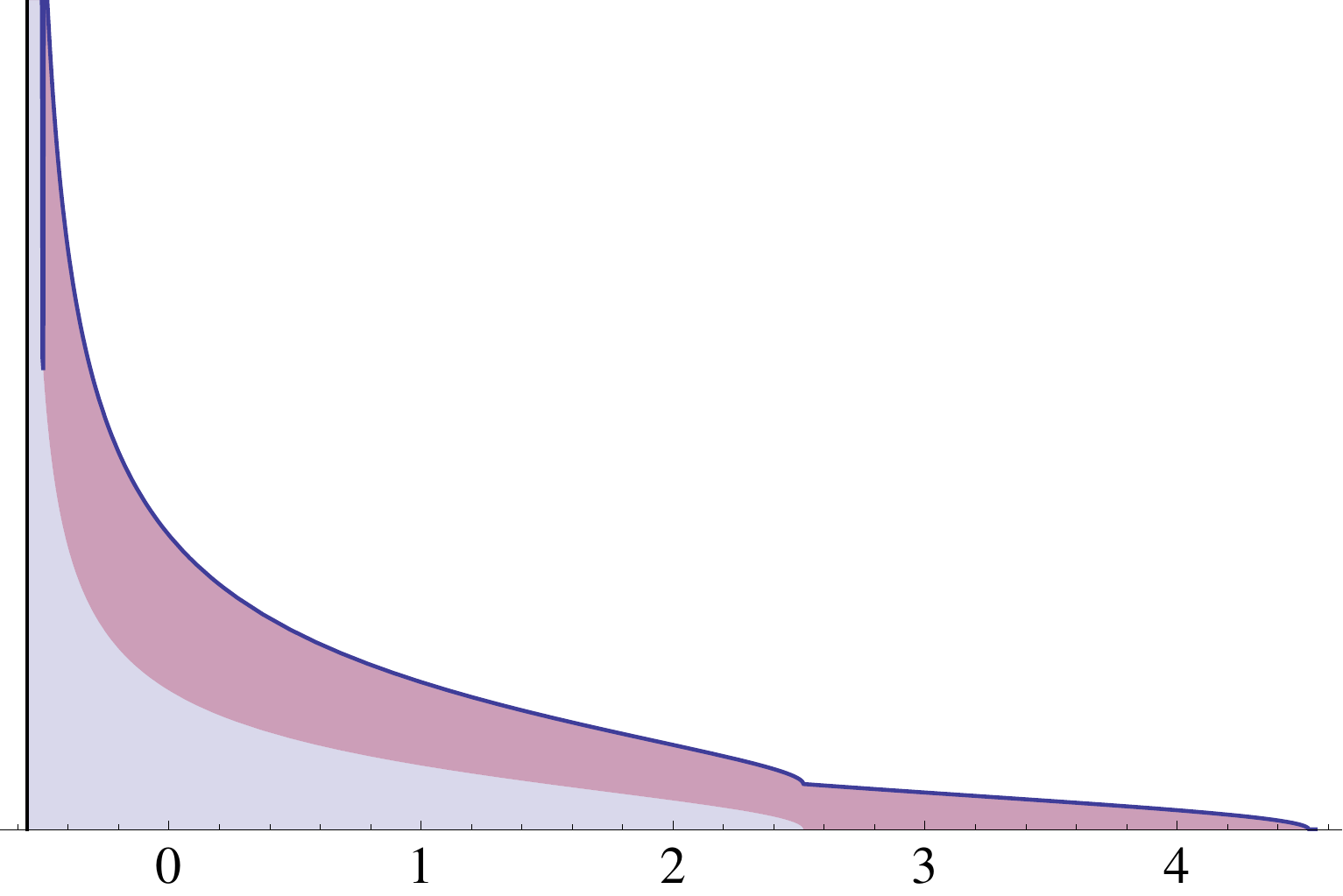}}\\
     \subfigure[$|W|=\ms$] {
        \includegraphics[width=0.43\textwidth]{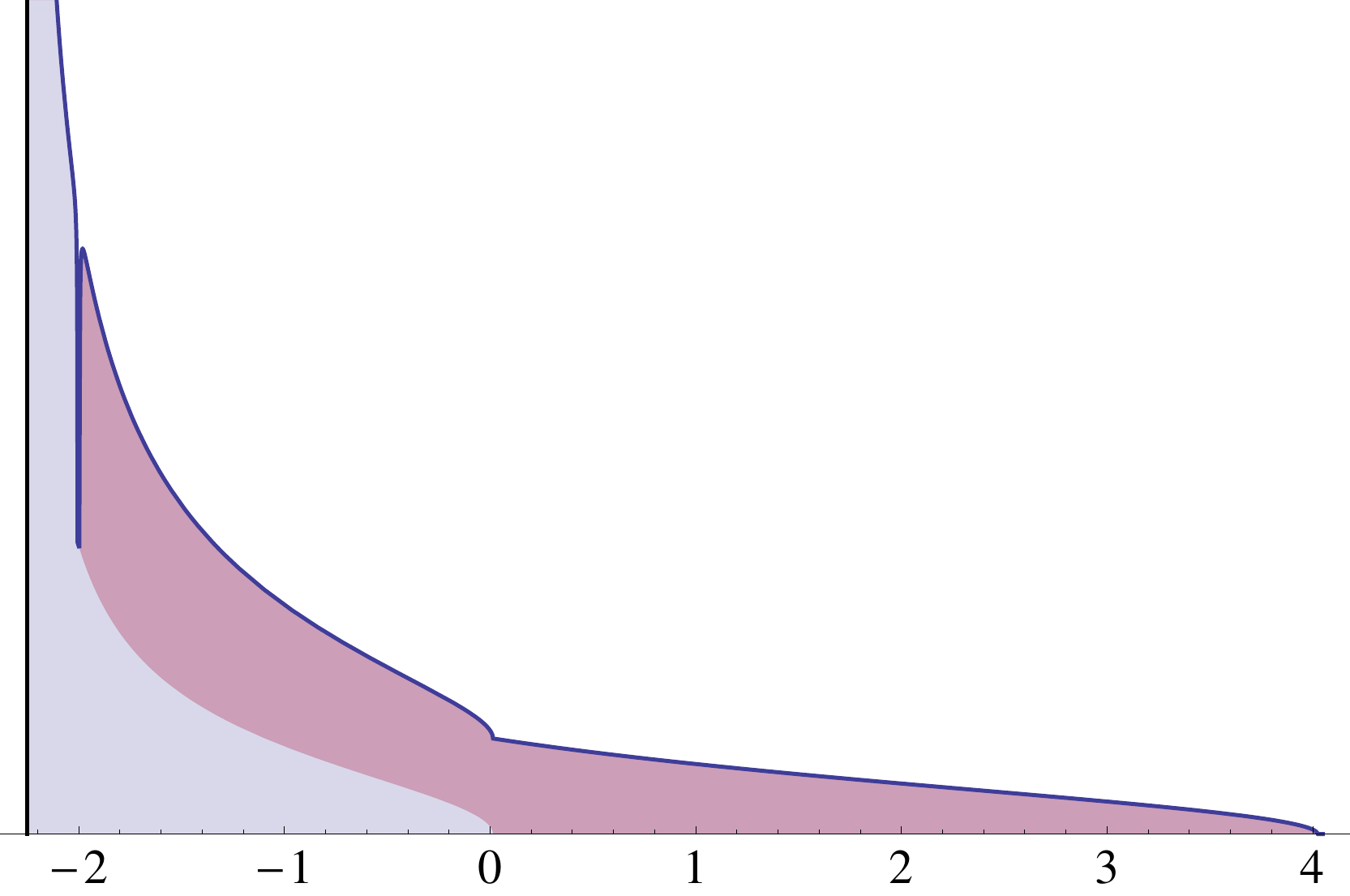}}
     \subfigure[$|W|=2~\ms$]{
        \includegraphics[width=0.43\textwidth]{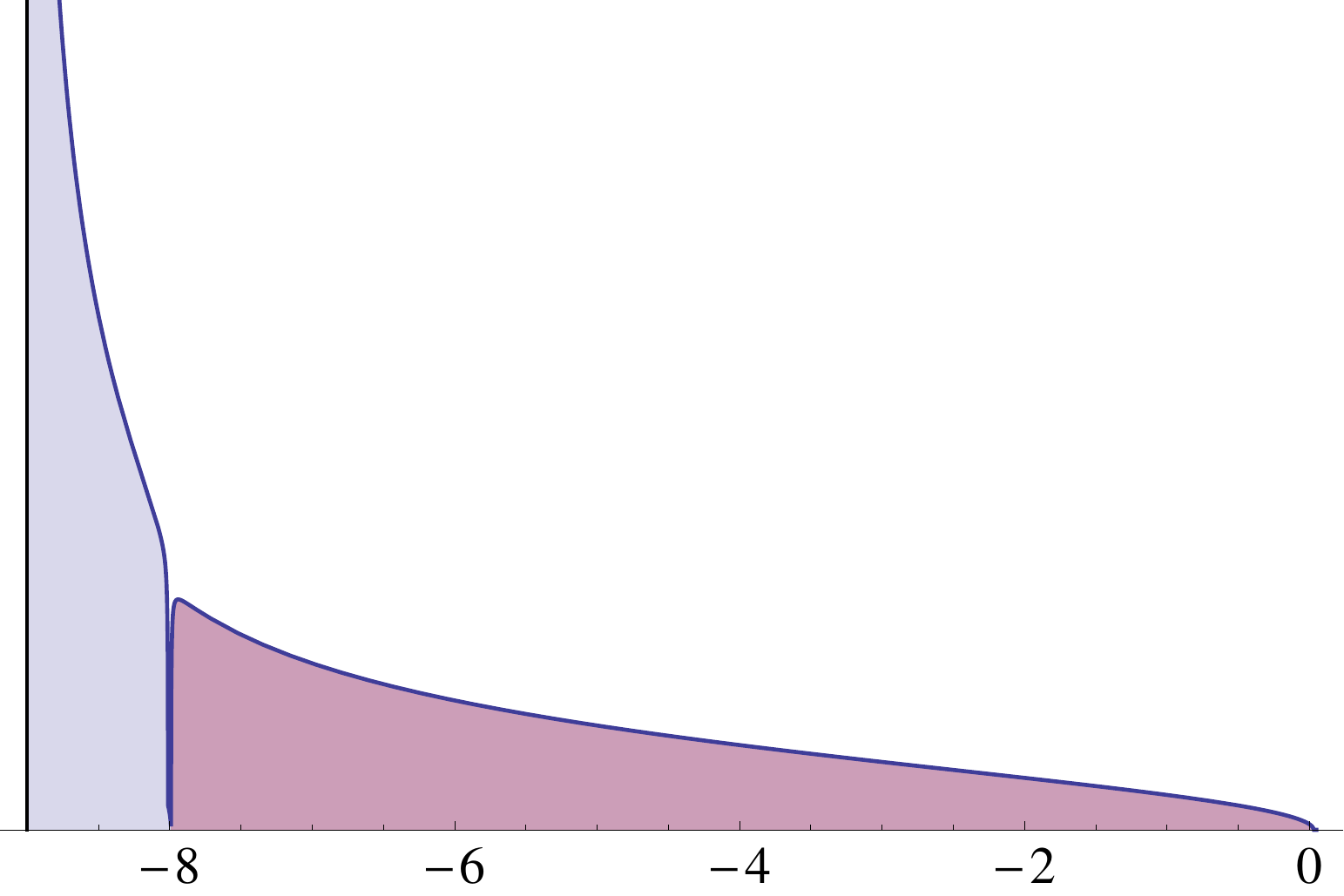}}\\
     \subfigure[$|W|=5~\ms$] {
        \includegraphics[width=0.43\textwidth]{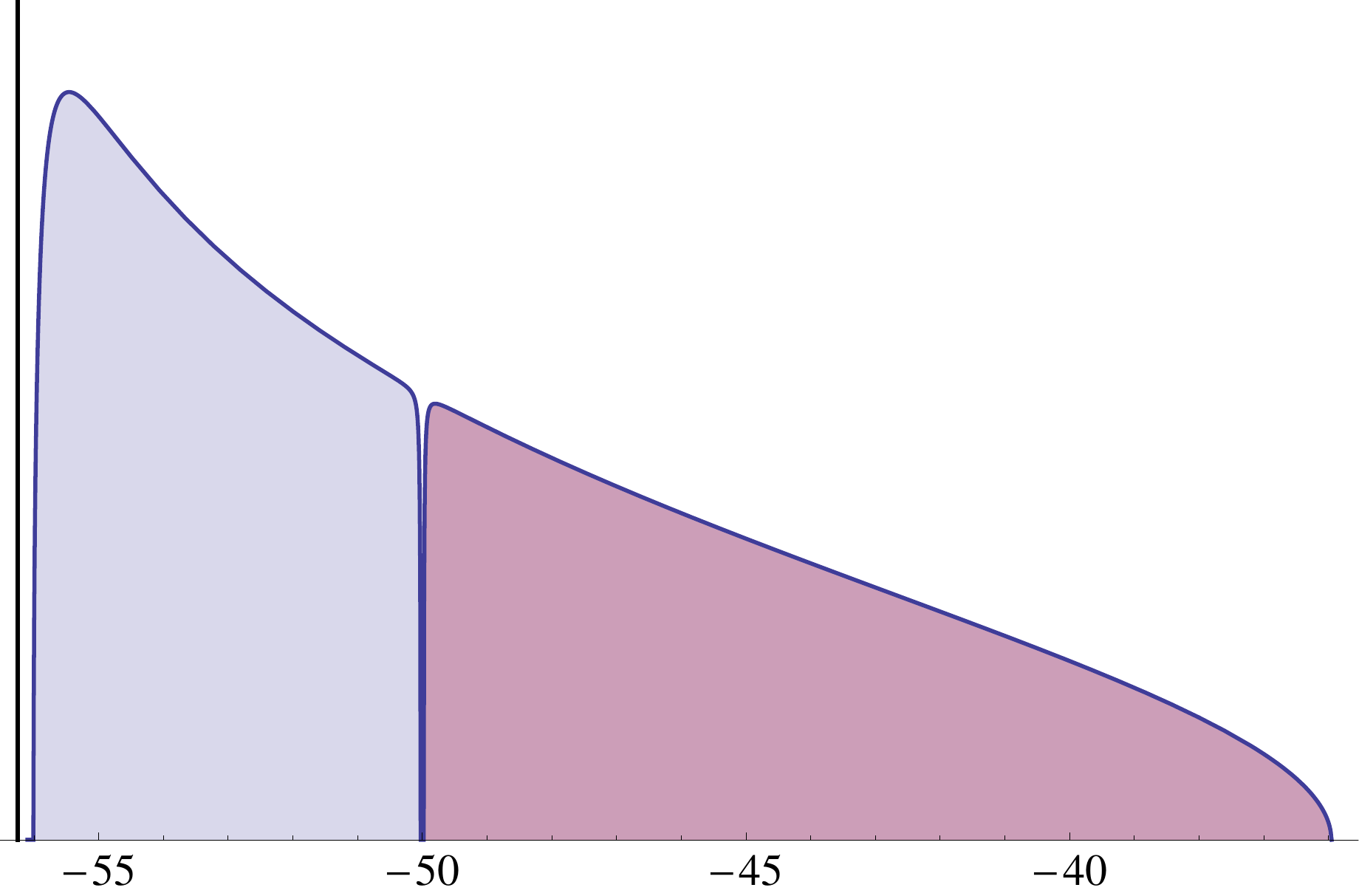}}
     \subfigure[$|W|=20~\ms$]{\label{fig:last}
        \includegraphics[width=0.43\textwidth]{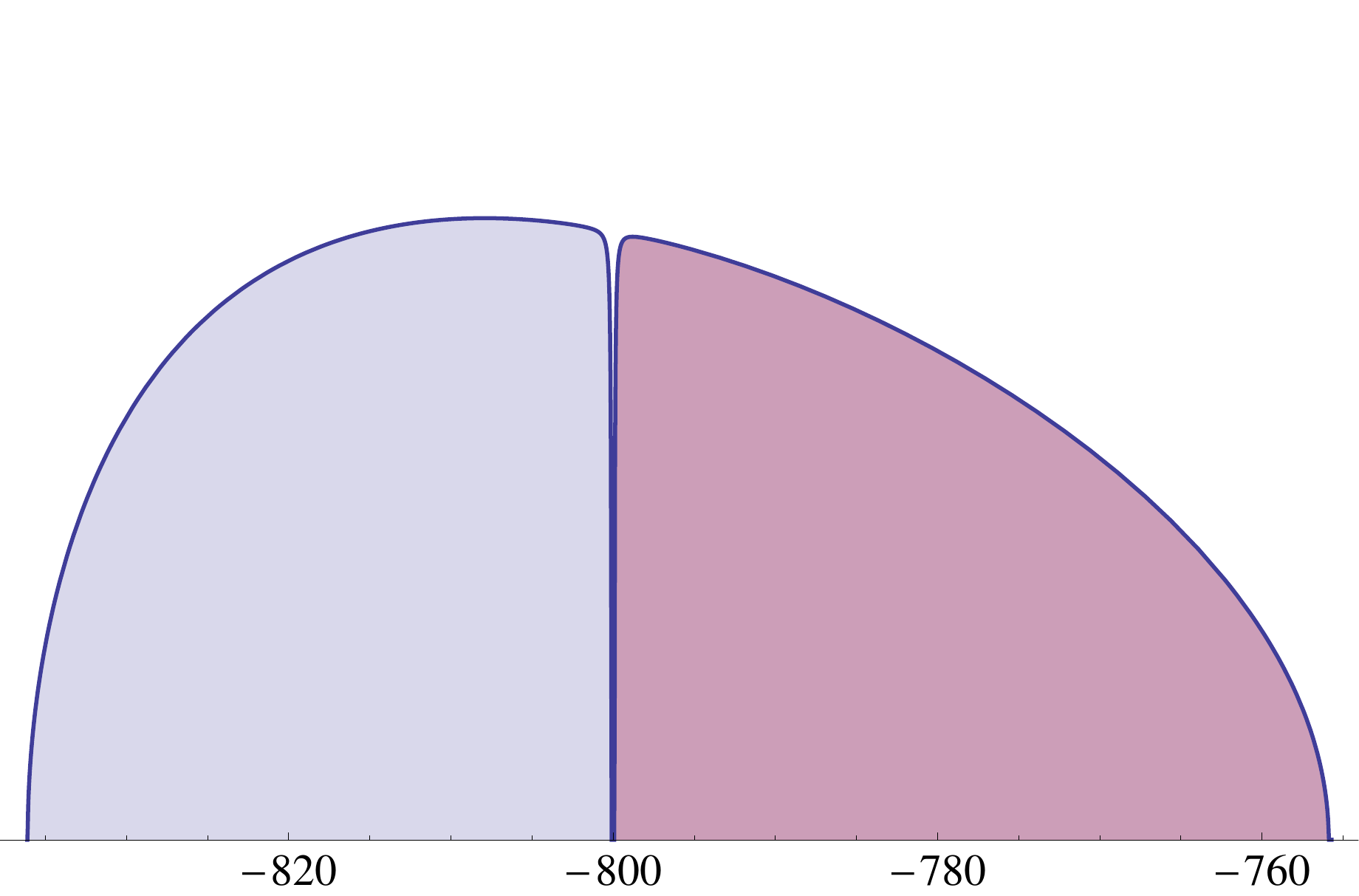}}
    \end{center}
    \caption{The mass spectra, in units of $\ms^2$, of generic supersymmetric AdS vacua with $N=100$ complex fields. The purple (darker) regions correspond to the contribution from the positive branch of \eqref{eq:omega2}, and the blue (lighter) regions correspond to the contribution from the negative branch. The black vertical line on the left side is at the BF bound $m^2_{\text{BF}} =-\tfrac94 |W|^2$ (not shown for plot \ref{fig:last}).}
   \label{fig:RandomAdS}
\end{figure}

The spectrum of the Hessian is given by a simple translation by $-2 |W|^2$ of the spectrum of ${\cal H}_Z$, and by equation \eqref{eq:BF}, the hard lower edge of equation \eqref{eq:omegamspectrum} corresponds to the BF bound for eigenvalues of the Hessian. Note that after the shift the negative branch has only negative support if $|W| > \ms$, and for $|W| > 2\,\ms$ all masses are tachyonic. Also, by equation \eqref{eq:omegampdf}, for $|W| \leq 4\,\ms$ the probability density $\rho_{\cal H}(m^2)$  has a square root divergence at the BF bound, which leads to a substantial amount of tachyons with masses close to the bound. We have plotted the typical
mass spectrum for a supersymmetric AdS vacuum for
the six different values $|W| = \tfrac{1}{100}\ms, \tfrac{1}{2}\ms, \ms, \, 2\,\ms, \, 5\,\ms, \, 20\,\ms$ in Figure \ref{fig:RandomAdS}.  One can see that the shape of the spectrum interpolates between the Wishart spectrum (cf.~Figure \ref{fig:Wishart}) for small $|W|$ and the Altland-Zirnbauer C$I$ spectrum (cf.~Figure \ref{fig:AZCI}) for large $|W|$.

Using equations \eqref{eq:Oplus} and \eqref{eq:Ominus}, we find that the total fraction of scalar fields that are tachyonic can be expressed in terms of the Wishart eigenvalue $\hm$ as
\be
f_{tachyons} = \frac{1}{2}\Big[ P\left( \hm \leq 4 |\hW|^2\right) + P\left(\hm \leq |\hW|^2\right) \Big] \, ,
\ee
which e.g.~for $|W|=\ms$ is around $80\%$.

The cleft in the Altland-Zirnbauer C$I$ spectrum translates into a similar cleft in the supersymmetric mass spectrum at $m^2 = -2|W|^2$. As in the case of the C$I$ ensemble, the cleft is a subleading effect in $1/N$, with a width that is inversely proportional to $N$. Moreover, as the probability distribution of $\omega_-$ goes to zero at the right edge of the Wishart spectrum at $\ha= \zeta_+ = 4 +  {\cal O}(1/N)$,
there is  a kink in the supersymmetric mass spectrum at $m^2 = 4\,\ms^2- 2|W|(\ms+|W|)$. We have verified these features through explicit numerical simulations, as detailed in the appendix.

\subsection{Fluctuations to positivity} \label{sec:fluctuations}
As we have just shown, typical supersymmetric vacua in supergravity have BF-allowed tachyons. In anti de Sitter space, these tachyons do not destabilize the vacuum, but in order to obtain a  background with a positive cosmological constant the AdS solution needs to be uplifted.   While it is not impossible that a suitably constructed uplift potential may stabilize several of the tachyons, requiring that the resulting de Sitter critical point is tachyon free is a significant restriction on string theory  model building. It is therefore interesting to consider the supersymmetric AdS vacua that are free of tachyons, or more generally have  masses only  above a certain  bound. This subset of solutions would appear more likely to result in metastable de Sitter vacua after uplifting.

In this section we compute the fraction of supersymmetric vacua with masses  above any given positive bound by mapping the problem of the positivity of the eigenvalues of the Hessian to that of fluctuations of the corresponding real, almost square Wishart matrix \cite{2011JSMTE..11..024B, Katzav:2010zz}. We present an analytic formula for the probability of such fluctuations as a function of the minimal mass squared, $|W|$, $\ms$, and $N$, and we derive the corresponding mass spectrum.

\subsubsection{The probability of positivity}

As we have discussed above, the scalar mass spectrum of the ensemble of supersymmetric vacua is determined in terms of the eigenvalues of a real Wishart matrix with $M=N+1$. At large $N$, the eigenvalue spectrum of the Wishart matrix is given by the Mar\v{c}enko-Pastur law, which is bounded by $\hm \in [\eta_-, \eta_+]$, as in equation \eqref{eq:wishart}. However, while the Mar\v{c}enko-Pastur law determines the \emph{typical} spectrum of eigenvalues, more rare configurations exist.  Here, we are interested in fluctuations of the smallest Wishart eigenvalue $\hm_{\text{min}}$ such that $\hm_{\text{min}} \geq 4|\hW|^2$. In this case the smallest eigenvalue of the Hessian ${\cal H}$ is given by
\be\label{eq:positivity}
m_{\text{min}}^2(\hm_{\rm min}) = \ms^2 \cdot \lp \hm_{\text{min}} -|\hW| \sqrt{\hm_{\text{min}}} - 2 |\hW|^2\rp \geq 0\,,
\ee
and the mass spectrum is tachyon free. Note that for $|W|\ge \ms$ these fluctuations require that all Wishart eigenvalues fluctuate past the generic right edge of the spectrum at $\eta_+ = 4 + {\cal O}(1/N)$.

The probability of a fluctuation to positivity is therefore
given by the probability that the smallest eigenvalue of the corresponding real Wishart matrix, with $M=N+1$, fluctuates to $\hm_{\text{min}} \ge 4|\hW|^2$.
In the case of real Wishart matrices with $M=N+1$ for which the entries of $A$ appearing in (\ref{Wishartdef}) are normally distributed, Edelman has obtained the {\it{exact}} distribution of the smallest eigenvalue, making no large $N$ approximation \cite{Edelman:1988:ECN:58846.58854}:
\be
\rho_{\hm_{\text{min}}}(x) = N^2 \,\chi_2^2(N^2 x) = \frac{N^2}{2} e^{-\tfrac12 N^2 x}\,. \label{eq:mumin}
\ee
Universality very plausibly leads to compatible results for more general distributions of the entries of $A$, once $N$ is sufficiently large.

The probability that the smallest eigenvalue $\hm_{\text{min}}$ is larger than some value $\zeta$ is then given by
\be\label{eq:min_distribution}
P(\hm_{\text{min}} \geq \zeta) = \int_\zeta^\infty dx \,\rho_{\hm_{\text{min}}}(x) = e^{-\tfrac12 N^2 \zeta} \, .
\ee
Thus, the probability that a random, supersymmetric AdS vacuum has no tachyons is
\be\label{eq:fluctuation_prob}
P(m^2 \geq 0) = \exp{(-2 N^2 |\hW|^2)} = {\rm exp}(-2N^2|W|^2 /\ms^{2})\,.
\ee
This is one of our main results.
Unless $|\hW| \lesssim {\cal O}(1/N)$, fluctuations to positivity are extremely unlikely at large $N$.

\subsubsection{The fluctuated spectrum}

The eigenvalue spectrum that results from a fluctuation to positivity can be computed using the Coulomb gas method as discussed in e.g.~\cite{Dean:2006wk, PhysRevE.77.041108}.
In \cite{Katzav:2010zz}, Katzav and Perez Castillo studied Wishart matrices for which all eigenvalues satisfy $\hm \geq \zeta \geq \eta_-$, finding the eigenvalue spectrum (for $\sigma = 1/\sqrt{N}$ and $M=N+1$)
\be\label{eq:fluctuatedMP1}
\rho_{MP}^{\text{fluc}}(\hm) = \frac{\left(\hm-\frac{1}{N}\sqrt{\frac{\zeta }{g(\zeta)}}\right) \sqrt{g(\zeta)-\hm}}{2 \pi  \hm \sqrt{\hm-\zeta}} \, ,
\ee
with
\be\label{eq:fuctuatedMP2}
g(\zeta)=\frac{4 (2+N (4+\zeta ))}{3 N} \sin^2 \ls\frac{\pi}{6} +\frac13 \text{arccot}\left[\sqrt{\frac{27 N \zeta }{(2+N (4+\zeta ))^3-27 N \zeta }}\right]\rs \,.
\ee
At the left, hard edge $\zeta$ of the fluctuated Wishart spectrum, the spectrum \eqref{eq:fluctuatedMP1} has a characteristic square root divergence. The right edge of the fluctuated Wishart spectrum is at $\hm_{\text{max}} = g(\zeta)$, which for $\zeta=4|\hW|^2$ is at $\hm_{\text{max}} = 4(1+|\hW|^2) + {\cal O}(1/N)$.

\begin{figure}
  \begin{center}
   \subfigure[$|W|=\tfrac{1}{2}~\ms$] {
        \includegraphics[width=0.45\textwidth]{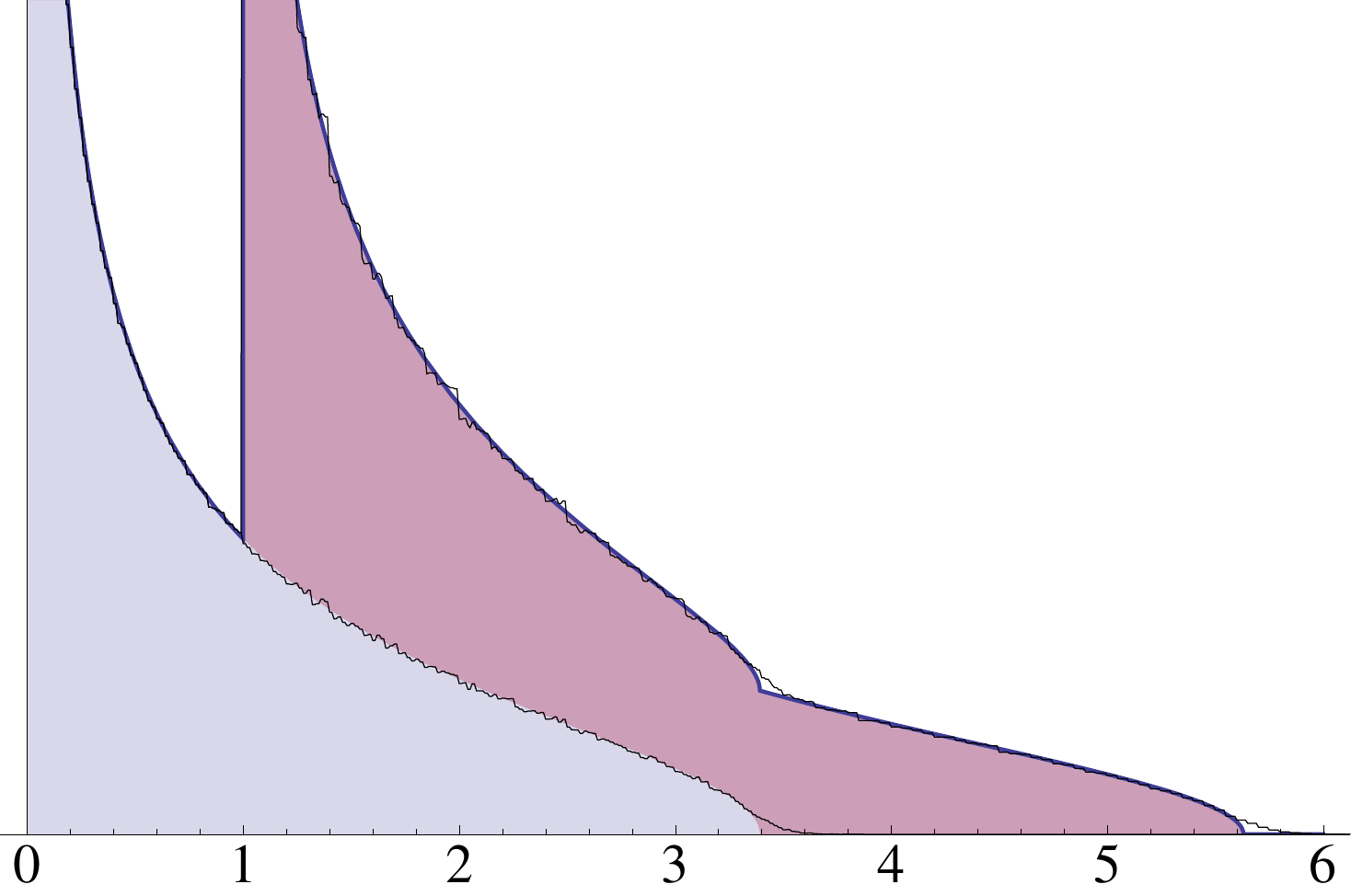}}
     \subfigure[$|W|=\ms$]{
        \includegraphics[width=0.45\textwidth]{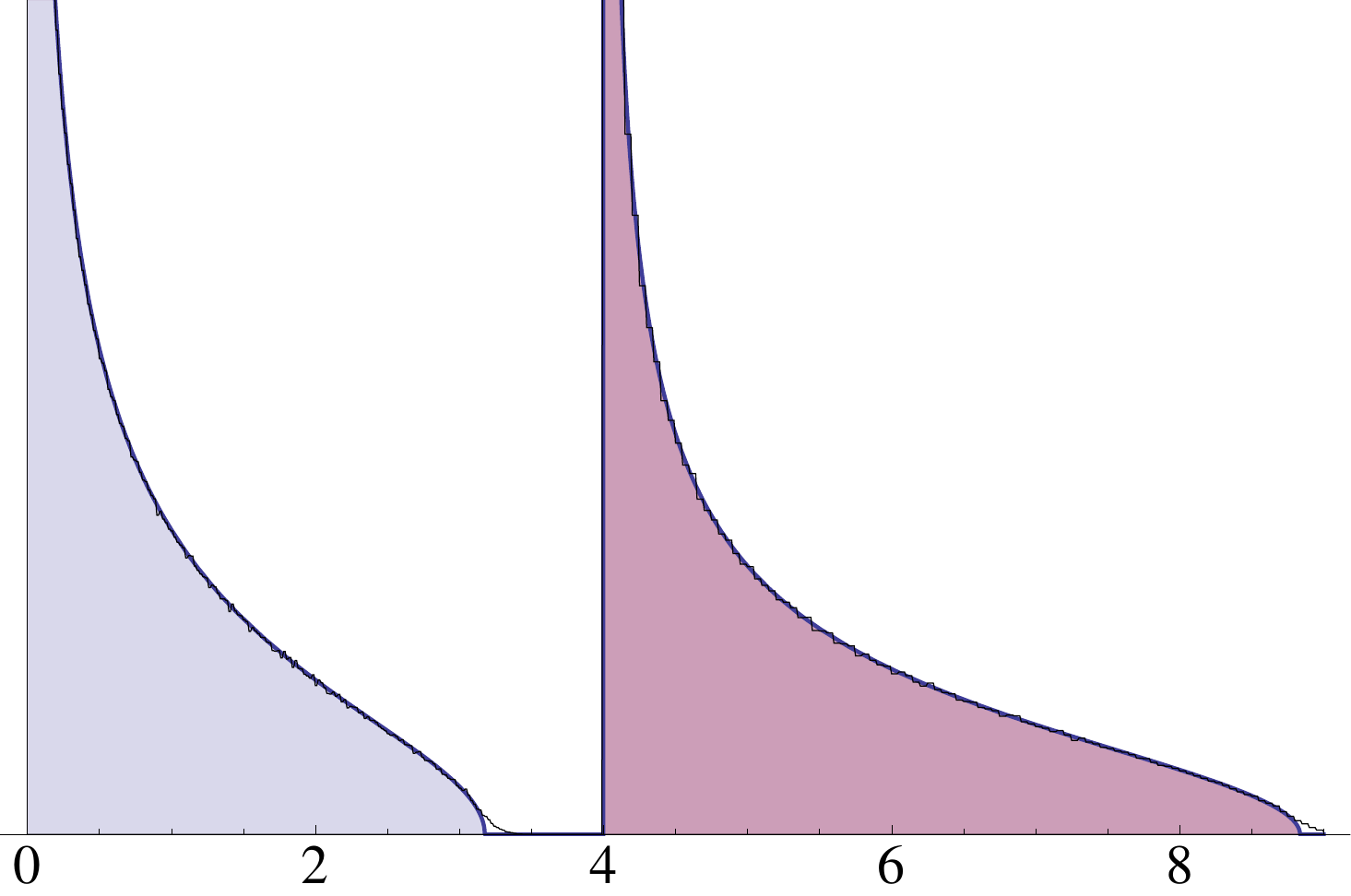}}
  \end{center}
  \caption{
  The mass spectra, in units of $\ms^2$, of supersymmetric AdS vacua with $N=100$ complex fields that have fluctuated to positivity. The purple (darker) regions correspond to the contribution from the positive branch of \eqref{eq:omega2}, and the blue (lighter) regions correspond to the contribution from the negative branch.  Blue curve: analytic result from equation \eqref{eq:fluctuatedH}.  Black line: equilibrium eigenvalue distribution from Metropolis simulation (see Appendix \ref{theappendix}).} \label{hessianmetropolis}
\end{figure}

As shown in \S\ref{sec:Hspectrum}, the spectrum of ${\cal H}_Z$ is given in terms of the  (now fluctuated)  spectrum of the corresponding Wishart matrix. By conditioning on $\hm \geq \zeta=4|\hW|^2$, the second term in \eqref{eq:Ominus} vanishes, so that the fluctuated spectrum of ${\cal H}_Z$ is given by
\be
\rho_{{\cal H}_Z}^{\text{fluc}}(\ha) = \frac{1}{2} \left[ \lp 1 + \tfrac{|\widehat W|}{\sqrt{4 \ha +|\widehat W|^2}} \rp \rho_{MP}^{\text{fluc}}(\hb_+) + \Theta(\ha) \lp 1 - \tfrac{|\widehat W|}{\sqrt{4\ha +|\widehat W|^2}} \rp \rho_{MP}^{\text{fluc}}(\hb_-)  \right] \, ,
\ee
where again $\hb_{\pm} = \frac{|\hW|^2}{2}+\ha \pm|\hW| \sqrt{\ha + \frac{|\hW|^2}{4}}$, and $\rho_{MP}^{\text{fluc}}(\hm)$ is given in \eqref{eq:fluctuatedMP1}, \eqref{eq:fuctuatedMP2}. From equation \eqref{Hess}, we note that upon including the shift, the probability density of the fluctuated scalar mass spectrum is
\be\label{eq:fluctuatedH}
\rho_{\cal H}^{\text{fluc}}(m^2) =  \rho_{{\cal H}_Z}^{\text{fluc}}(m^2 + 2|W|^2) \, .
\ee
The two branches of eigenvalues of ${\cal H}$ have support for
\be
\zeta \pm |\hW| \sqrt{\zeta} -2|\hW|^2 \leq \frac{m_\pm^2}{\ms^2} \leq 4+2|\hW| \lp |\hW| \pm \sqrt{1 +|\hW|^2} \rp \, .
\ee
Note that for $\zeta=4|\hW|^2$, this means that the branches do not overlap when $|\hW| \geq \tfrac{2}{\sqrt{5}}$. The correlation between the two branches leads to two distinct peaks in the mass spectrum, at each of which the spectrum exhibits a square-root divergence.
In Figure \ref{hessianmetropolis} we have plotted the fluctuated spectrum for $\zeta = 4|\hW|^2$ and for two different values of $|\hW|$.

\subsection{The distribution of tachyon-free vacua} \label{distributionsection}

Let us conclude this section by considering the full distribution of supersymmetric vacua, and the constrained distribution of tachyon-free supersymmetric vacua, in random supergravity.  The number of vacua can be computed as \cite{Denef:2004ze}
\be
N_{vac.} = \int~d{\cal P}[W, F, Z, \ldots]~\delta^{(2N)}(F)~|\det \partial_A F_B| \, ,
\ee
where the capital indices $A, B$ run over both holomorphic and antiholomorphic indices. The measure $d{\cal P}$ specifies the prior distribution of the fully covariant Taylor coefficients of the superpotential, and as in the rest of this paper, we will make the assumption that the integration variables  are independent and identically distributed, while allowing for different standard deviations  for $|W|$, $|F_a|$ and $|Z_{ab}|$. In particular, this means that the prior measure is separable and, with some abuse of notation, we write $d{\cal P}[W, F, Z] = d{\cal P}[W] d{\cal P}[F]  d{\cal P}[Z]$. With these assumptions we have
\bea
N_{vac.} &=& C_N~\int~d{\cal P}[W]~\int d{\cal P}[F] ~\int d{\cal P}[Z] ~\delta^{(2N)}(F)~|\det \partial_A F_B| \nonumber \\
&=& C_N~\int~d{\cal P}[W]~\int d{\cal P}[F] ~\int d{\cal P}[Z] ~\delta^{(2N)}(F)~\left|~\det \left(
\begin{array}{c c}
\bW \delta_{a \bb}  & Z_{ab} \\
\bZ_{\bar a \bar b} & W \delta_{\bar a b}
\end{array}
\right)~\right|
\, . \qquad
\eea
The integral over $Z_{ab}$ can be written as an integral over the eigenvalues and eigenvectors of the matrix ${\cal M}$ in equation \eqref{eq:M}. By applying the $2N \times 2N$ unitary transformation of equation \eqref{U}, it is easy to see that the above integral simplifies to
\be
N_{vac.} = \tilde C_N~\int~d{\cal P}[W]~\int d{\cal P}[F] ~\int d{\cal P}[\lambda_1, \ldots, \lambda_N ]~\delta^{(2N)}(F) ~\prod_{a=1}^N \left| \lambda_a^2 - |W|^2\right| \, ,
\ee
where the integration over eigenvectors of ${\cal M}$ is suppressed. The  Jacobian factor $\prod_{a=1}^N \left| \lambda_a^2 - |W|^2\right|$  ensures that each vacuum counts with unit weight in the integral by canceling the corresponding inverse factor arising from the delta function.

The frequency of tachyon-free supersymmetric vacua is given by
\bea
f &=& \frac{N_{vac.}(m^2_{min}\geq0)}{N_{vac.} }  \\
&=& \frac{\int~d{\cal P}[W]~\int d{\cal P}[F] ~\int d{\cal P}[\lambda_1, \ldots, \lambda_N ]~\delta^{(2N)}(F) ~\prod_{a=1}^N \left| \lambda_a^2 - |W|^2\right|~e^{-2N^2 |W|^2/\ms^2}
}{\int~d{\cal P}[W]~\int d{\cal P}[F] ~\int d{\cal P}[\lambda_1, \ldots, \lambda_N ]~\delta^{(2N)}(F)\,\prod_{a=1}^N \left| \lambda_a^2 - |W|^2\right|} \, , \nn
\eea
where $m_{susy}$ is determined as in the discussion around equation \eqref{eq:Zhat}.

Of particular interest is the region in which $\lambda_1 \gtrsim |W|$, or equivalently $\ms \gtrsim N\,|W|$.
In this region there is no exponential suppression of tachyon-free vacua,
and  in fact tachyon-free vacua are quite common.  For $d{\cal P}[W]$ being roughly uniform in the complex $W$ plane, a fraction $1/N^2$ of all supersymmetric vacua, and {\it most} tachyon-free supersymmetric vacua, occur in this region. The spectrum in this more stable region consists of two overlapping branches
approximately following the Mar\v{c}enko-Pastur law, as illustrated in Figure \ref{fig:first}. However, the absence of BF-allowed tachyons in a supersymmetric solution does not guarantee the existence of an uplift to a metastable de Sitter vacuum, as we will now discuss.

\section{Instabilities of Uplifted Vacua} \label{implications}

Our results so far concern the prevalence of tachyons in supersymmetric vacua, but the question of primary physical interest is the likelihood of tachyonic instabilities in a realistic vacuum with positive cosmological constant.  The `uplift' paradigm advanced in \cite{KKLT} suggests dividing the problem into two parts: ascertaining the mass spectrum in a supersymmetric vacuum, and incorporating stabilizing or destabilizing effects of the supersymmetry-breaking energy.  In \S\ref{sec:spectrum} we have given a comprehensive treatment of the first point for ${\cal N}=1$ supergravities that are `random' in the sense of \cite{Marsh:2011aa}, and we now turn to quantifying the effects of uplifting.

Instabilities of random critical points in various models of a supergravity landscape have been studied in a number of works, which we briefly review for completeness.
The original analysis of \cite{Aazami:2005jf} modeled the mass matrix at a critical point as the sum of a real Wigner matrix and a positive definite diagonal matrix,
and quantified the probability of instability in terms of the size of the Wigner component.  More recently, \cite{Chen:2011ac} worked with a similar model of the mass matrix, motivated by constructions in type IIA string theory.  The stability of de Sitter vacua resulting from \emph{spontaneous} breaking of supersymmetry by an F-term was the subject of \cite{Denef:2004cf,Marsh:2011aa}. In those works the supersymmetry breaking was taken to
result from the same $W$ and $K$ that gave rise to the moduli potential, rather than through an uplifting due to supersymmetry breaking in another sector.  In general, explicit uplifting from a supersymmetric solution to a de Sitter solution can arise by integrating out the supersymmetry breaking sector of a theory with spontaneously broken supersymmetry, but not all conceivable uplifts need to be of this form.

Given the mass spectrum in a supersymmetric vacuum, the crucial question is whether the source of uplifting introduces new instabilities, cures existing instabilities, or leaves the mass matrix unchanged.
A very simple and presumably unrealistic picture that is often invoked in the literature is a `rigid' uplift, in which the cosmological constant is increased but the mass matrix is unmodified.
In this case, our counting of tachyons in supersymmetric vacua translates directly to the de Sitter critical points resulting from uplifting.
In particular, in view of equation (\ref{eq:fluctuation_prob}), the probability that a rigidly-uplifted de Sitter configuration is metastable is
\be
P(m^2 \geq 0) = \exp{\left(-2 N^2 |W|^2/m_{susy}^2\right)}\,,
\ee so that metastability is extremely rare for $|W| \gg m_{susy}/N$.

\subsection{Wigner uplift} \label{wu}
While modeling the uplift as rigid serves as a simple starting point, models in which the uplift potential has a nontrivial field dependence are more general and appear more likely to arise from supersymmetry breaking in string compactifications. For a general function $V_{up}$ of $N$ complex scalar fields, the contribution to the Hessian
is given by
\be \label{vupeq}
{\cal H}_{up} = \left(
\begin{array}{c c}
\partial^2_{a \bb} V_{up} & \partial^2_{a b} V_{up} \\
\partial^2_{\bar a \bb} V_{up} & \partial^2_{\bar a b} V_{up}
\end{array}
\right) \, ,
\ee
where $\partial^2_{a \bb} V_{up} = \left(  \partial^2_{\bar a b} V_{up} \right)^*$ is an $N \times N$ Hermitian matrix and $\partial^2_{a b} V_{up}$ is an $N \times N$ complex symmetric matrix.
We will take the entries of $\partial^2_{a \bb} V_{up}$ to be distributed as the i.i.d.\ entries of an $N \times N$ Hermitian matrix with the natural scale $V_{up} \approx 3|W|^2$, and the entries of $\partial^2_{a b} V_{up}$ to be distributed as the i.i.d.\ entries of an $N \times N$ complex symmetric matrix of the same scale.

We note that ${\cal H}_{up}$ is `time-reversal' symmetric,
\be
T {\cal H}_{up} T^{\dagger} = {\cal H}_{up}^* \, ,
\ee
under the unitary time-reversal operator $T= + T^\top = \sigma_1$ acting on the $N\times N$ block matrices. Since this is the only symmetry of ${\cal H}_{up}$, we identify ${\cal H}_{up}$ as a representation of the A$I$ symmetry class \cite{Dyson:1962es, Altland:1997zz}, more commonly known as the Gaussian orthogonal ensemble.  Thus, although we have taken $V_{up}$ to be a function of complex scalars, ${\cal H}_{up}$ is in the symmetry class\footnote{In our numerical treatment we have simulated matrices of the form \eqref{vupeq} and found perfect agreement with the Gaussian orthogonal ensemble, strongly suggesting that these matrices are in fact in the same universality class.} of a real symmetric matrix, just as in the model of \cite{Aazami:2005jf}.

Supersymmetric vacua in which $m_{susy} \ll N|W|$ have many tachyons in AdS, and as a Wigner uplift potential of the form (\ref{vupeq}) serves to disperse the supersymmetric spectrum, these vacua will predominantly uplift to unstable critical points in Minkowski or de Sitter space. The probability of obtaining special configurations that enjoy a larger likelihood of uplifting to a metastable vacuum was discussed in \S\ref{sec:fluctuations}, and here we simply note that in order to produce a vacuum once the uplift potential is taken into account, the spectrum of the AdS vacuum typically has to fluctuate far enough so that the left edge of the spectrum is well above zero, which is extremely unlikely in general.

A much more interesting region occurs for $|\hW|\lesssim 1/N$, in which case the AdS vacuum  is typically tachyon free, and the uplift potential will only produce a minor perturbation in the spectrum.
We find it convenient to parametrize the standard deviation of the entries of ${\cal H}_{up}$ as $\sigma_{{\cal H}_{up}}= c_{2}^{2} m_{susy}^2/N^{2}$, for a dimensionless constant $c_2$, and we parametrize  $|\hW|$ as  $|\hW| = c_{1} /N$. Note that the natural scale $V_{up} \approx 3|W|^2$ corresponds to $c_1 \approx c_2$. Specializing to a basis in which the supersymmetric Hessian is diagonal, ${\cal H} = {\rm diag}(m^2_l)$, it is easy to compute the eigenvalues of the total Hessian,  ${\cal H}_{tot}={\cal H}+{\cal H}_{up}$, to second order in perturbation theory:
\be
(m^2_{tot})_{l}=m^2_{l }+({\cal H}_{up})_{ll}-\sum_{i\ne l}\frac{|({\cal H}_{up})_{il}|^{2}}{(m^2_{i}-m^2_{l})}+\ldots,\label{eq:evexpansion}
\ee
where the ellipsis includes terms of higher order in ${\cal H}_{up}$.
The typical size of the smallest eigenvalue can be estimated from \eqref{eq:positivity}, using the fact that from (\ref{eq:mumin}), the mean of the smallest eigenvalue of the associated Wishart matrix is $2/N^2$:
\bea
\langle(m^2_{tot})_{min}\rangle &=& m_{susy}^2 \left( \frac{2}{N^{2}}-\sqrt{2}\frac{|\hW|}{N}-2 |\hW|^{2}-\sqrt{2} \frac{\sigma_{{\cal H}_{up}}}{\ms^2} \right)+\ldots  \nonumber \\
&\approx& \frac{\ms^2}{N^2} \left(2- \alpha\right)
  \, ,
\eea
where $\alpha \equiv \sqrt{2}c_{1} + 2  c_{1}^{2} + \sqrt{2} c_{2}^{2}$, and the quadratic corrections have been neglected. Thus, to leading order a typical de Sitter critical point arising from uplifting remains stable if $c_1, c_2 \ll 1$.

It remains to check that higher order corrections are subleading. For small $|\hW|$, the eigenvalues $m^2_{l}$ are nearly doubly degenerate, and the two smallest eigenvalues are split by $\langle m^2_{2}-m^2_{1} \rangle = 2 \sqrt{2} |\hW|\ms^2/N  = 2\sqrt{2} c_{1} \ms^2/N^{2}$.  For $c_{1}\ll 1$ this is the smallest difference of eigenvalues in the spectrum, and the expectation value of the quadratic correction to the smallest eigenvalue can be approximated by (cf. \eqref{eq:evexpansion})
\be
\Big\langle \sum_{i\ne 1}\frac{|({\cal H}_{up})_{i1}|^{2}}{(m^2_{i}-m^2_{1})} \Big\rangle\approx \frac{c_{2}^{4}}{N^{4}} \frac{\ms^2}{2 \sqrt{2} c_{1}/N^{2}}\approx \frac{c_{2}^{4}}{c_{1} }~\frac{\ms^2}{2 \sqrt{2}N^2}\,,
\ee
so that higher order corrections are indeed subleading for $c_1 \approx c_2$.

\begin{figure}
  \begin{center}
    \includegraphics[width=0.5\textwidth]{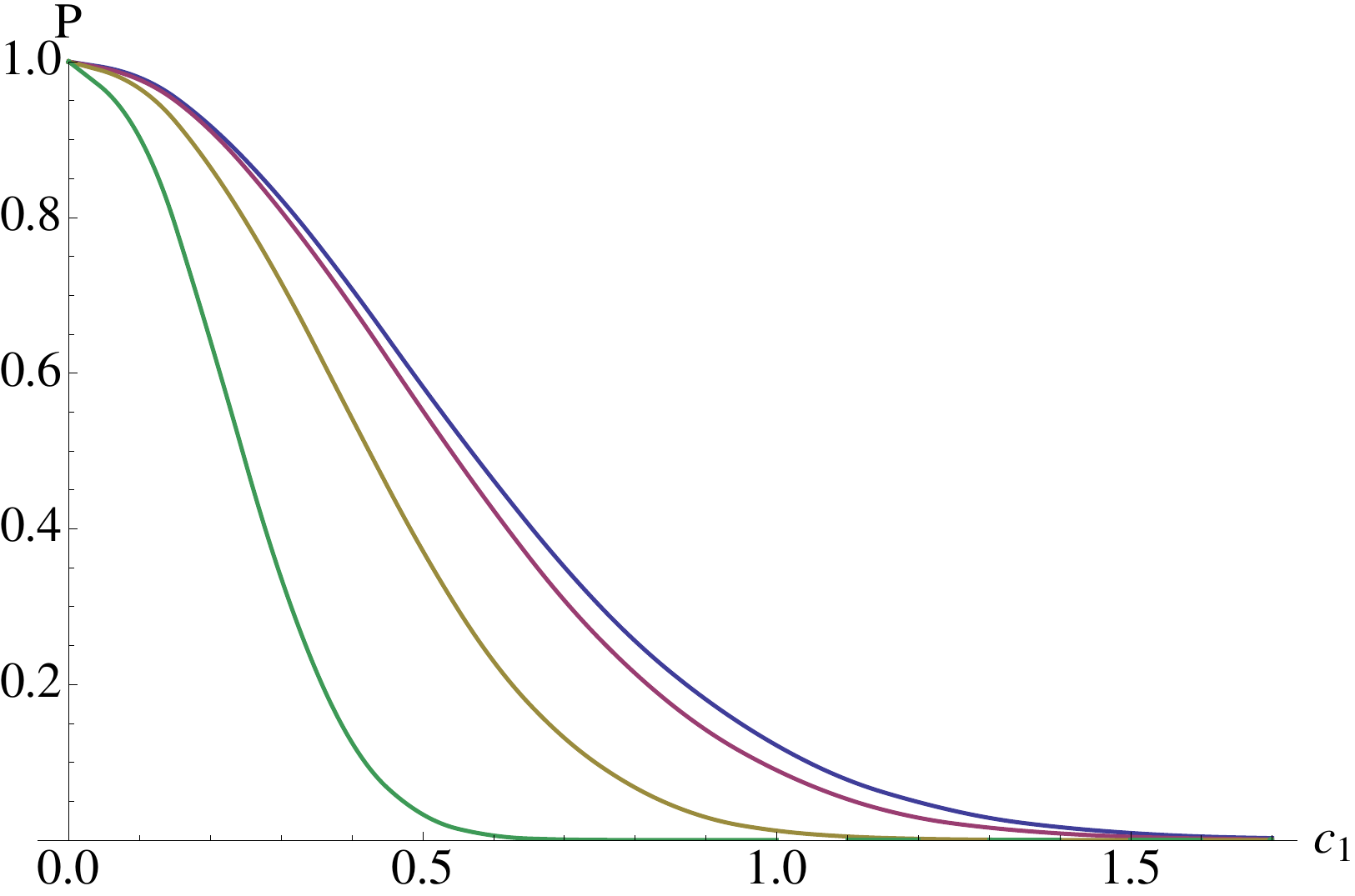}
  \end{center}
  \caption{The probability of the absence of tachyons in $\mathcal{H}_{tot}=\mathcal{H}+\mathcal{H}_{up}$ for $|\hW|=c_{1}/N$, versus $c_{1}$ for $N$=10 and $c_{2}=2\sqrt{3}c_{1}, \, \sqrt{3}c_{1},\,c_1,\, 0$ (from left to right).}\label{fig:HplusHup}
\end{figure}

We have verified these analytic result through extensive numerical simulations. The probability of the absence of tachyons in the mass spectrum of $\mathcal{H}_{tot}=\mathcal{H}+\mathcal{H}_{up}$ for different values of $c_1$ and $c_2$ is shown in Figure \ref{fig:HplusHup}.

\subsection{D-terms}
\label{sec:D}
Although it is plausible that many sources of supersymmetry breaking and uplifting will introduce instabilities, as described in the preceding section, we now point out that the presence of D-terms can contribute a stabilizing effect.  A comprehensive treatment of D-terms in random  supergravity is beyond the scope of this work, and we will content ourselves with a few observations.

For an $\mathcal{N}=1$ supergravity with the product gauge group $G=G_1 \times G_2 \times \ldots \times G_n$, the scalar potential contains the additional term
\be
V_D=\frac12\lp\text{Re} f\rp^{-1\,ij}D_i D_j\,,
\ee
where $f_{ij}$ denotes the holomorphic gauge kinetic function, and $D_i$ is the D-term for the gauge group $G_i$, which (for field configurations with $W\ne 0$) is given by \cite{Nilles:1983ge}
\be
\label{eq:GeneralDTerm}
D_i= \frac{i}{W} D_aW\, X^{a}_i \, ,
\ee
where $\epsilon^i X_i^a$ is the variation of the field $\phi^a$ under an infinitesimal gauge transformation $A^i\rightarrow A^i+d\epsilon^i$. Here, $X^a_i$ denotes the components of the Killing vector of the corresponding isometry of the Kähler geometry, which for linearly realized gauge symmetries is given by $X_i^a = -i T_i \phi^a$,  where we have suppressed the gauge indices for the generator $T_i$ and for $\phi^a$.

A supersymmetric solution satisfies $D_aW=0$, so that it follows from \eqref{eq:GeneralDTerm} that $V_D$ vanishes at any supersymmetric critical point. For this reason D-terms cannot be used to uplift an F-flat minimum.
However, the D-term potential can change the masses of the scalar fields.
In particular, since $V_D$ is positive semidefinite and vanishes at a supersymmetric vacuum, it can only contribute positively to the eigenvalues of the Hessian, and may therefore decrease the likelihood of tachyons. However, the presence of D-terms restricts the contribution to the Hessian matrix arising from $V_F$ as follows: since the D-terms are real, $(\partial_a D_i)^* = \partial_{\bar a} D_i$, which for supersymmetric solutions implies that\footnote{This relation may also be obtained directly from the Killing equation.} $Z_{ab} X^b_i = - \delta_{a \bb} W \bar X^{\bb}_i$, where we have specialized to $K_{a \bb} = \delta_{a \bb}$ at the point in question. This consistency condition can be written as an eigenvalue equation for the $2N$ component  vectors $\tilde X^{\pm}_i = (\bar X^{\bar a}_i , \pm X^a_i)^\top$ as
\be
\widetilde {\cal M} \tilde X^{\pm}_i = \mp|W| \tilde X^{\pm}_i \, ,
\ee
where
\be\label{eq:Mtilde}
\widetilde {\cal M} = \left(
      \begin{array}{cc}
        0 & Z_{a b}~e^{- i \vartheta_W} \\
        \bar{Z}_{\bar a  \bar b}~e^{ i \vartheta_W}& 0 \\
      \end{array}
    \right) \,
\ee
has the same spectrum as the matrix ${\cal M}$ of equation \eqref{eq:M}, and was identified in \cite{Denef:2004cf} as the matrix governing the critical point equation of spontaneously broken F-term supergravity. Here $\vartheta_W$ denotes the phase of the superpotential, $W = e^{i \vartheta_W} |W|$.  It follows from equation \eqref{eq:omega} that the contribution to the Hessian from the  F-term potential will have eigenvalues $0$ and $-2|W|^2$ in the directions $\tilde X^{\pm}_i$.

Using $D_aW=0$ and redefining the fields  $A^i$ to set  $\text{Re} f_{ij}=\delta_{ij}$ at the point in question, we find the D-term contribution to the Hessian in a supersymmetric vacuum:
\be
\mathcal{H}_D = \left(
                  \begin{array}{cc}
                    \bar{X}_{ia} \, X^i_{~\bar{b}} & - \bar{X}_{ia} \,  \bar{X}^i_{~b} \\
                    - X_{i\bar{a}}\, X^i_{~\bar{b}} & X_{i\bar{a}} \, \bar X^i_{~b} \\
                  \end{array}
                \right) = \sum_{i=1}^n \left(
                  \begin{array}{c}
                    \bar{X}_{ia} \\
                    - X_{i \bar{a}} \\
                  \end{array}
                \right) \left( X^i_{~\bar{b}}\,,\,  -  \bar{X}^i_{~b} \right) =  \sum_{i=1}^n \tilde X^-_i~\tilde X^{-  i \dagger} \,,
\ee
where the indices $a,b$ are lowered by the metric and the index $i$ is raised by the gauge kinetic function. The D-term contribution to the Hessian for any given $G_i$ is thus of rank one, with a nonvanishing contribution only in the directions specified by $\tilde X^-_i$.

We  conclude that for $n \ll N$, D-terms do not substantially alter the results described in \S\ref{wu}, but do lead to the existence of special directions along which the eigenvalues of the Hessian are given by  $0$ and $-2|W|^2 + || \tilde X^-_i ||^2$.

In cases where $n\gtrsim N$, contributions from the D-terms may have a substantial impact on the mass spectrum and the number of tachyons.  In order for the D-terms to affect the stability of the Hessian arising from the F-term potential, a substantial fraction of the $N$ scalar fields have to carry charges under the gauge groups $G_i$.  Understanding the extent to which this condition is met in well-motivated compactifications is an interesting problem that we will not address in this work.

\section{Conclusions} \label{conclusions}

We have computed the scalar mass spectrum in a supersymmetric vacuum of a general four-dimensional ${\cal N}=1$ supergravity theory, with the Kähler potential
and superpotential taken to be random functions of $N \gg 1$ complex scalar fields.
By relating the spectrum of the Hessian matrix to the spectrum of an associated Wishart matrix,
we showed that a fraction $f = {\rm exp}(-2N^2|W|^2/\ms^2)$ of AdS vacua are tachyon-free.
Then, using results from the Coulomb gas formulation of random matrix theory, we obtained the scalar mass spectrum resulting from a fluctuation to positivity.
We performed extensive numerical cross-checks of our analytic expressions for the fluctuation probability and mass spectrum, with perfect agreement.

A clear implication of our results is that uplifting a supersymmetric vacuum with $|W| \gtrsim \ms/N$ through the addition of positive-energy sources is overwhelmingly likely to lead to an unstable de Sitter critical point, not a metastable de Sitter vacuum.
The likelihood of uplifting to a de Sitter vacuum can be increased by making the supersymmetric mass scale $\ms$ large compared to
$|W|$.  For $\ms \gtrsim N|W|$ and $N\gg 1$, the probability that uplifting a supersymmetric AdS vacuum leads to a metastable de Sitter vacuum can be vastly larger than the probability that a general de Sitter critical point with spontaneously broken supersymmetry is metastable, cf.\ \cite{Marsh:2011aa}.

Our results are applicable when the superpotential and K\"ahler potential are accurately described as random functions of $N$ complex fields.  Determining the prevalence of tachyons in AdS vacua in more general theories with multiple sectors or special structures in $K$ and $W$ is an interesting problem for the future.

\subsection*{Acknowledgements}
We are grateful to M.~Douglas, A.~LeClair, J.~P.~Sethna, and S.~Zelditch for helpful discussions.
This research was supported by the NSF under grant PHY-0757868.
D.M.~gratefully acknowledges support from the Gålö foundation. The research of L.M.~was supported by an NSF CAREER Award.
The work of T.W.~was supported by a Research Fellowship (Grant number WR 166/1-1) of the German Research Foundation (DFG).

\appendix

\section{Numerical results} \label{theappendix}
In \S\ref{spectrum} we obtained an analytic expression for the supersymmetric mass spectrum ${\cal H}_Z$ and computed the probability of a fluctuation to positivity.
The simplest and most direct check of these results, the direct diagonalization of a large sample of Wishart matrices, is a computationally intensive task: most of the parameter space of interest corresponds to large fluctuations that are extremely rare (i.e. $ |\hW|\sim 1$ and $N\gg 1$ in \eqref{eq:fluctuation_prob}).  A more efficient approach
utilizes Monte Carlo methods to obtain both the fluctuation probability and the spectrum of fluctuated eigenvalues. In the following, we present two numerical techniques to obtain these quantities.

To begin, we will review a few elements of the Coulomb gas formulation of random matrix theory \cite{Dyson:1962es, Dyson:Brownian}, which underlies the analytic results presented in \S\ref{sec:fluctuations}, but also serves as an efficient starting point for simulations \cite{PhysRevE.77.041108}.

Recall that the joint probability density function of the eigenvalues of a
Wishart matrix is given by \eqref{eq:Wishart}. The integral over this joint pdf can be interpreted as the partition function of a gas of interacting particles
undergoing Brownian motion
in one-dimensional space at finite temperature.
The eigenvalues are then interpreted as the positions of particles that are confined by a linear background potential and repel each other with a force that is inversely proportional to the distance between any pair of particles.\footnote{This force may be interpreted as a two-dimensional Coulomb interaction between equally charged particles moving in one-dimensional space, hence the name of the formalism.}
The Hamiltonian of this $N$ particle system  is then given by
\be\label{hamiltonian}
H=\frac{1}{\sigma^{2}} \sum_{a=1}^N \mu_a \ - 2\sum_{a < b}^N{\rm{ln}}|\mu_a - \mu_b| - \xi\,\sum_{a=1}^N {\rm{ln}}\,\mu_a\, ,
\ee
where we impose the constraint $\mu_a>0$, which corresponds to a hard wall at $\mu=0$. The temperature of the gas is given by $2/\beta$ --- recall that $\beta=1,2$ for real and complex Wishart matrices, respectively.

The Coulomb gas formalism
suggests that techniques and intuition from statistical mechanics can be useful in the study of random matrices.
In fact, by rewriting the partition function in terms of the empirical eigenvalue density $\rho(\mu) = {\tiny \frac{1}{N}} \sum_{i=1}^N \delta(\mu - \mu_i)$, and taking the continuum limit, the partition function can be written as a path integral over the scalar field $\rho$, and saddle-point evaluation can be used to find the equilibrium distribution \eqref{eq:wishart}.

Moreover, saddle-point evaluation can be used to find the probability density  of the eigenvalues of \emph{constrained} configurations, such as those in which the smallest eigenvalue is larger than some cutoff $\zeta$, as well as the probability of obtaining such a configuration. In \S\ref{sec:fluctuations}, we used solutions of the fluctuated Wishart spectrum obtained in \cite{Katzav:2010zz} to find the spectrum of fluctuated eigenvalues of the Hessian matrix analytically.  In this appendix, we confirm  these findings by numerical simulation of the $N$-particle Coulomb gas.

\begin{figure}
  \begin{center}
    \includegraphics[width=0.6\textwidth]{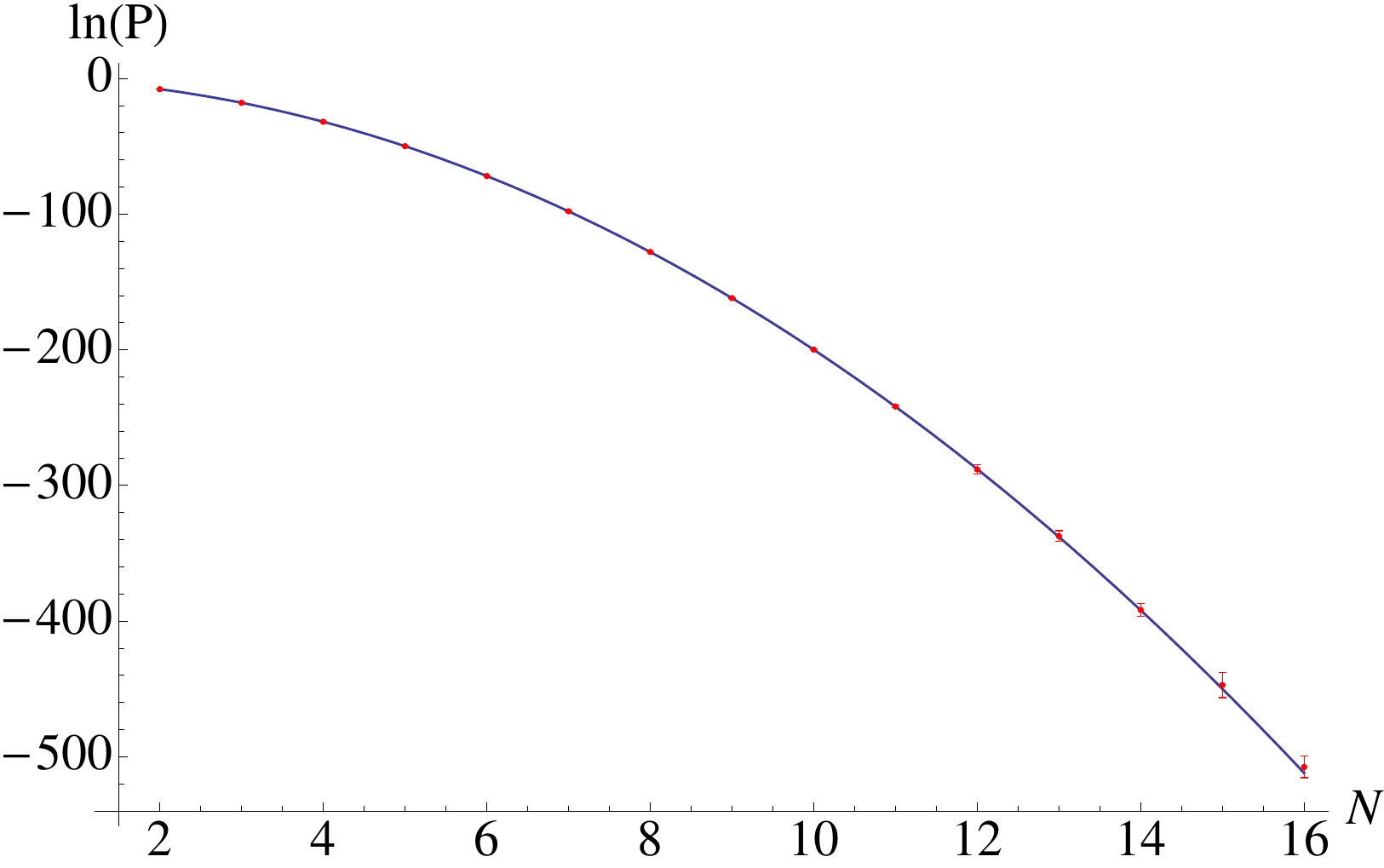}
  \end{center}
  \caption{The probability of a fluctuation to positivity of $\mathcal{H}_{Z}$ for $|\hW|=1$, versus the number of complex scalar fields $N$. The curve is the analytic result given in equation (\ref{eq:fluctuation_prob}), and the dots correspond to $6\times10^{10}$ Monte Carlo samplings with $3\sigma$ error bars.}\label{montecarlofluctuation}
\end{figure}

The probability of obtaining a fluctuated solution is given by
\be
P(\mu_{\text{min}} \geq \zeta)=\frac{Z_{N}(\zeta)}{Z_{N}(-\infty)}\, ,
\ee
where the constrained partition function $Z_{N}(\zeta)$ is given by
\be\label{partition}
Z_{N}(\zeta)=\int_{\zeta}^{\infty}d\mu_{1}\int_{\zeta}^{\infty}d\mu_{2} \ldots \int_{\zeta}^{\infty}d\mu_{N}\,\exp\left(-\frac{\beta}{2} H(\mu )\right).
\ee
As in \cite{PhysRevE.77.041108}, the integral in \eqref{partition} can be computed numerically via Monte Carlo sampling by considering the average $\langle \cdot \rangle_{\zeta}$, where the $\mu_a>\zeta$ are randomly chosen from a uniform distribution. This leads to the probability of the smallest eigenvalue
\be
P(\mu_{\text{min}} \geq \zeta)=\lim_{\Lambda\rightarrow \infty} \left(\frac{\Lambda-\zeta}{2\Lambda}\right)^{N} \frac{\left\langle e^{-\beta H(\mu)/2}\right\rangle_{\zeta}}{\left\langle e^{-\beta H(\mu)/2} \right\rangle_{-\Lambda}}\,,
\ee
where the prefactor captures the average spacing between the sampling points. We used $6 \times 10^{10}$ Monte Carlo samplings to evaluate the fluctuation probability numerically and obtained excellent agreement with the analytical result, as shown in Figure \ref{montecarlofluctuation}.

While the method of evaluating the partition function directly via Monte Carlo integration captures the fluctuation probability, it does not provide the probability density function of the eigenvalues. In particular, we are interested in the pdf of a fluctuated spectrum, as computed in \eqref{eq:fluctuatedMP1}. However, these fluctuations are extremely rare for the parameters of interest, and are essentially inaccessible by direct diagonalization of matrices. Instead, we will take advantage of the Coulomb gas picture and numerically simulate the equilibrium distribution.

A popular algorithm to thermalize classical particles in an arbitrary potential is the Metropolis algorithm \cite{annealing2}. This algorithm shifts one particle at a time to obtain the equilibrium distribution in the following way: (1) select one particle, (2) obtain the energy cost for performing a random step, (3) if the energy cost is negative, keep the step. Otherwise, keep the step with probability $e^{-\beta \Delta E/2}$. In order to avoid artifacts from the simulation, the size of each step should be small compared to the typical particle spacing.
When starting with an arbitrary particle distribution there is a typical number of steps $n_{\text{thermalize}}$ required after which the initial correlations decay and the gas reaches thermal equilibrium. In our simulations we chose a fixed number of steps $n_{\text{steps}}\gg n_{\text{thermalize}}$ to avoid correlations from the initial configuration but still observe any thermal fluctuations. While implementing the Metropolis algorithm for the Hamiltonian in \eqref{hamiltonian} leads to a generic random distribution of eigenvalues, we can further constrain the Hamiltonian by introducing a hard wall at $\mu=\zeta$, allowing us to access the eigenvalue distribution for large fluctuations.  The result from Metropolis simulation, along with the analytic result, appears in Figure \ref{hessianmetropolis}.

\bibliographystyle{modifiedJHEP}
\bibliography{refs}
\end{document}